\documentclass[
reprint,
superscriptaddress,
amsmath,amssymb,
prx,
]{revtex4-2}

\usepackage{graphicx}% Include figure files
\usepackage{dcolumn}% Align table columns on decimal point
\usepackage{bm}% bold math
\usepackage{hyperref}% add hypertext capabilities
\usepackage[dvipsnames]{xcolor}
\usepackage{soul}
\usepackage{ulem}
\usepackage[left]{lineno}
\usepackage{braket}

\definecolor{dark-green}{RGB}{0, 128, 0}

\begin{document}
%\preprint{APS/123-QED}
\title{Edge-dependent anomalous topology in synthetic photonic lattices\\ subject to discrete step walks}
%Edge-dependent edge states: anomalous topology in discrete step walks studied in a synthetic photonic lattice

\author{Rabih~El~Sokhen}
\affiliation{Univ. Lille, CNRS, UMR 8523 -- PhLAM -- Physique des Lasers Atomes et Mol\'ecules, F-59000 Lille, France}

\author{\'Alvaro~G\'omez-Le\'on}
\email{a.gomez.leon@csic.es}
\affiliation{Institute of Fundamental Physics IFF-CSIC, Calle Serrano 113b, 28006 Madrid, Spain}

\author{Albert~F.~Adiyatullin}
\altaffiliation[Present address: ]{Quandela, 7 Rue Léonard de Vinci, 91300 Massy, France}
\affiliation{Univ. Lille, CNRS, UMR 8523 -- PhLAM -- Physique des Lasers Atomes et Mol\'ecules, F-59000 Lille, France}

\author{Stéphane~Randoux}
\affiliation{Univ. Lille, CNRS, UMR 8523 -- PhLAM -- Physique des Lasers Atomes et Mol\'ecules, F-59000 Lille, France}

\author{Pierre~Delplace}
\affiliation{ENS de Lyon, CNRS, Laboratoire de Physique (UMR CNRS 5672), F-69342 Lyon, France}

\author{Alberto~Amo}
\email{alberto.amo-garcia@univ-lille.fr}
\affiliation{Univ. Lille, CNRS, UMR 8523 -- PhLAM -- Physique des Lasers Atomes et Mol\'ecules, F-59000 Lille, France}

\date{\today}

\begin{abstract}
Anomalous topological phases, where edge states coexist with topologically trivial Chern bands, can only appear in periodically driven lattices. When the driving is smooth and continuous, the bulk-edge correspondence is guaranteed by the existence of a bulk invariant known as the winding number. However, in lattices subject to periodic discrete-step walks the existence of edge states does not only depend on bulk invariants but also on the boundary. This is a consequence of the absence of an intrinsic time-dependence or micromotion in discrete-step walks. We report the observation of edge states and a simultaneous measurement of the bulk invariants in anomalous topological phases in a two-dimensional synthetic photonic lattice subject to discrete-step walks. The lattice is implemented using time multiplexing of light pulses in two coupled fibre rings, in which one of the dimensions displays real space dynamics and the other one is parametric. The presence of edge states is inherent to the periodic driving and depends on the properties of the boundary in the implemented two-band model with zero Chern number. We provide a suitable expression for the topological invariants whose calculation does not rely on micromotion dynamics.
\end{abstract}

\maketitle

\section{Introduction}
Systems with topological properties are changing the way we understand phases of matter and physical phenomena. Their mathematical characterization, purely based on topological arguments, has allowed to find a plethora of new phases and to discover a large number of practical applications~\cite{Hasan2010,von-Klitzing2019,Li2014,ZhaoGuoXiaWang2015-NanoPhotApp}.
In addition, the extension to the realm of non-equilibrium or dissipative systems has increased the variety of topological phases and their possible uses~\cite{Gong2018}.

A particularly interesting and fruitful example of this are Floquet topological phases~\cite{Lindner2011,Gomez-Leon2013}, see Ref.~\cite{harper_topology_2020} for a review. They appear in systems with enriched topology due to the coupling to an external driving field that is periodic in time. In the high frequency regime, i.e., when the frequency of the driving field dominates over all the other energy scales, the system can be successfully described by a stroboscopic effective Hamiltonian. In this regime, the driving field serves as a tuning knob of the system's external parameters and can be used to engineer static pseudomagnetic fields~\cite{Rechtsman2013b,Flaschner2016}.
In contrast, when the frequency is reduced, stroboscopic dynamics is not enough to describe the system, and one needs to explicitly consider the evolution dynamics within a single driving period, also known as the micromotion. 
In this regime, the topology of Floquet systems becomes fundamentally different from that of static ones, and one can find anomalous topological phases with chiral edge states in systems with topologically trivial bands~\cite{Kitagawa2010,Kitagawa2012,Rudner2013,AGL2014,Nathan2015,peng_experimental_2016, cheng_observation_2019, afzal_realization_2020, Wintersperger2020, AGL-Floquet-2023, zhang_tuning_2023, braun_real-space_2023}.

Anomalous phases have also been found in systems with a discrete-step time evolution, known as quantum walks~\cite{Kitagawa2012,Asboth2013, barkhofen_measuring_2017,Maczewsky2017, Mukherjee2017a, Bellec2017, Bisianov2019}.
One of their particular features is the absence of an intrinsic time-coordinate due to their discrete time evolution. 
This implies that the topological characterization of Floquet phases with a Hamiltonian description, which requires an explicit time-coordinate~\cite{Rudner2013, Nathan2015}, is not suitable for quantum walks. 
Recent theory works have shown~\cite{Cedzich2018,Mochizuki2020,Grudka2023} that quantum walks result in a richer topological phase diagram due to their discrete time evolution, and that not only the bulk topology, but also the boundaries play a crucial role in the existence of chiral edge states~\cite{Bessho2022}. A direct consequence is that one can study a time-discretized version of the anomalous Floquet topological phase, characterized by bulk invariants given by generalized winding numbers~\cite{Rudner2013}, and show that, at the same time, the chiral edge states can be removed by choosing a suitable set of edge unitary evolution operators at the boundary. Hence, the bulk invariant is still well-defined, but it cannot predict the number of edge states for arbitrary boundaries, notably when the edge unitaries are topologically non-trivial~\cite{Bessho2022}.

Experimentally, Floquet topological phases in discrete step walks have been realized in photonic lattices~\cite{Kitagawa2012,Rechtsman2013b,barkhofen_measuring_2017, Maczewsky2017, Mukherjee2017a, Bellec2017, Bisianov2019, Mukherjee2020, afzal_realization_2020, Mukherjee2021}. 
However, the characterisation of anomalous phases in these experimental systems has been constrained either to chiral symmetric lattices in one-dimension~\cite{Kitagawa2012, barkhofen_measuring_2017, Bellec2017, Bisianov2019} or to two-dimensional lattices in the limit of a continuous time evolution~\cite{Maczewsky2017, Mukherjee2017a} employing the invariants discussed by Rudner and co-workers~\cite{Rudner2013, Nathan2015}, which do not take into account the crucial discrete-step aspect of the time evolution.

In this work, we simultaneously investigate the bulk and edge properties of anomalous topological phases in discrete step walks in a two-dimensional synthetic photonic lattice. The lattice is implemented using time multiplexing of light pulses in two coupled fibre rings.
In this system, one of the dimensions displays real space dynamics and the other one is defined by an external phase applied to the rings.
We focus on the relation between the bulk invariants and the presence of topological edge states, and demonstrate important differences with anomalous topological phases in Floquet systems with a Hamiltonian description (i.e. under continuous time evolution). In particular, we show that the specific choice of the boundaries affects the existence of edge states.
Our results are well described by suitably defined winding numbers which we directly measure in the experiment, and open the door to the engineering of extrinsic topological phases~\cite{Bessho2022}.
In these phases, the number of topological edge modes can be varied through the appropriate design of unitary operators at the edges of the lattice. The implementation of such configuration could be useful to switch on and off edge transport via local modifications. 

%%%%%%%%%%%%%%%%%%%%%%%%%%%%%%%%%%%%
\begin{figure}[t]
\includegraphics[width=0.8\columnwidth]{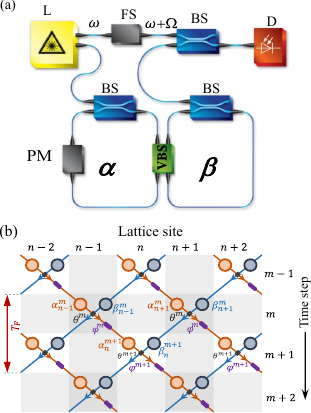}
\caption{\label{fig1} 
(a) Scheme of the double ring system with injection and reference laser L, beamsplitters BS, variable beamsplitter VBS, frequency shifter FS, phase modulator PM and photodiode D. (b) Synthetic split step lattice.
}
\end{figure}
%%%%%%%%%%%%%%%%%%%%%%%%%%%%%%%%%%%%%%%%

%%%%%%%%%%%%%%%%%%%%%%%%%%%%%%%%%%%%
\begin{figure}[t]
\includegraphics[width=\columnwidth]{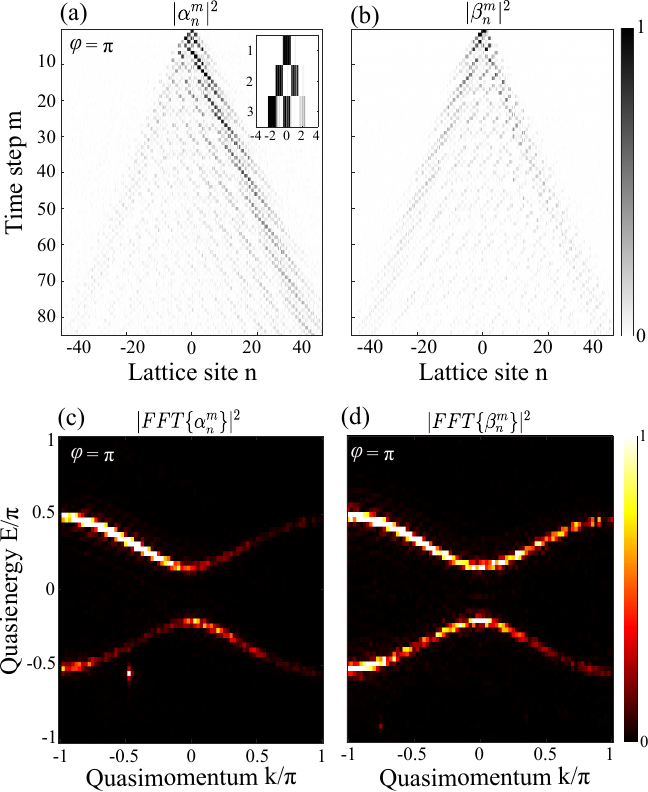}
\caption{\label{fig:spatiotemporal} 
(a), (b) Spatiotemporal dynamics measured in the $\alpha$ (a) and $\beta$ (b) rings for $\theta_1=0.2\pi$, $\theta_2=0.3\pi$ and $\varphi=\pi$. (c), (d) Measured bands extracted from the square of the Fourier transform of the even time steps of (a), (b), respectively.
}
\end{figure}
%%%%%%%%%%%%%%%%%%%%%%%%%%%%%%%%%%%%%%%%

\section{Experimental implementation of a discrete step walk}

To study discrete step walks we use time-multiplexed lattices made with coupled optical fiber loops.
This type of synthetic lattice has been used in different configurations as a test-bed to study, among others, the formation of solitons in parity-time symmetric lattices~\cite{Wimmer2015,Muniz2019}, Bloch oscillations~\cite{Wimmer2015a}, anomalous transport~\cite{Wimmer2015}, artificial gauge fields~\cite{Chalabi2019}, the non-hermitian skin effect~\cite{Weidemann2020}, superfluidity~\cite{Wimmer2021}, and Floquet winding bands~\cite{Adiyatullin2023}.
Importantly, in this setup it is possible to simultaneously access the bulk and edge properties.
The bulk can be studied from the direct measurement of the eigenvectors of the Floquet operator~\cite{Lechevalier2021} and via the measurement of anomalous transport~\cite{Wimmer2017}, which give access to the Berry curvature of the Floquet bands over the whole first Brillouin zone. 
The implementation of physical edges in the setup can be done via the onsite control of the splitting ratios in the step evolution.

In our experimental configuration the two fibre rings of 45.34~m and 44.79~m in length are coupled via a variable beamsplitter (VBS), sketched in Fig.~\ref{fig1}(a). A short squared pulse of about 1~ns at a wavelength of 1550nm is injected in the $\alpha$ ring and evolves in the rings following a split step walk each time it reaches the beamsplitter. The small difference in length between the rings allows encoding the lattice position information in the time of arrival of pulses at the extraction port at each round trip. This round trip time is the unit of step time evolution.
Er-doped amplifiers in the rings compensate for the insertion, extraction and absorption losses allowing the circulation of pulses for many round trips. Figure~\ref{fig:spatiotemporal}(a), (b) displays the measured step evolution in the $\alpha$ and $\beta$ rings when a pulse is injected at a single site in the $\alpha$ ring.

The dynamics of the amplitude and phase of light pulses in the rings can be mapped into a coherent step evolution in the one-dimensional synthetic lattice depicted in Fig.~\ref{fig1}(b), governed by the following equations~\cite{Bisianov2019,Weidemann2020}:

\begin{eqnarray} \label{Eq:step}
\alpha_n^{m+1} &=& \left(\cos\theta_m\alpha_{n-1}^m + i\sin\theta_m\beta_{n-1}^m\right) e^{i\varphi_m} 
\nonumber \\
\beta_n^{m+1} &=& i\sin\theta_m\alpha_{n+1}^m + \cos\theta_m\beta_{n+1}^m
\label{eq:EOM1},
\end{eqnarray}

\noindent with $\alpha_n^{m}$ and $\beta_n^{m}$ being the complex amplitude of the light pulses in the left and right rings, respectively, at lattice position $n$ and round trip time step $m$. The splitting amplitude at the variable beamsplitter is $\theta_m$, which can be controlled electronically and alternates between two values $\theta_1$ and $\theta_2$ at odd and even steps. To get the second synthetic dimension, a phase modulator adds a controlled phase $\varphi_m$ to the $\alpha$ ring with a value that alternates between $\varphi$ and $-\varphi$ at odd and even steps. The lattice sites $n$ provide a spatial dimension along which dynamics can take place with an associated conjugated momentum $k$, while the external phase $\varphi$ acts as a generalised quasimomentum resulting in a second \textit{parametric} dimension ($\varphi \in (-\pi, \pi]$).

The lattice system has two spatial sublattices corresponding to the $\alpha$ and $\beta$ rings (red and blue circles in Fig.~\ref{fig1}(b)) and a Floquet period of two time steps. Applying the Floquet-Bloch ansatz $(\alpha^{m}_{n},\beta^{m}_{n})^{\dagger}=(\alpha,\beta)^{\dagger}e^{iEm/2}e^{ikn/2}$ to Eq.~\ref{Eq:step} we find two bands of eigenvalues $E_{\pm}(k,\varphi) =\pm\arccos[\cos\theta_{1}\cos\theta_{2}\cos(k)- \sin\theta_{1}\sin\theta_{2}\cos(\varphi)]$. The factor $1/2$ in the exponents of the ansatz take into account the two sites spatial periodicity and the period of two time steps  in Eq.~\ref{Eq:step}. Due to the periodicity in time of the lattice, $E$ is defined modulo $2\pi$, and the two bands are separated by two gaps centered around $E=0$ and $E=\pi$ in the first Floquet Brillouin zone.
%****************************************
%****************************************
%****************************************
\begin{figure*}[t]
\includegraphics[width=\textwidth]{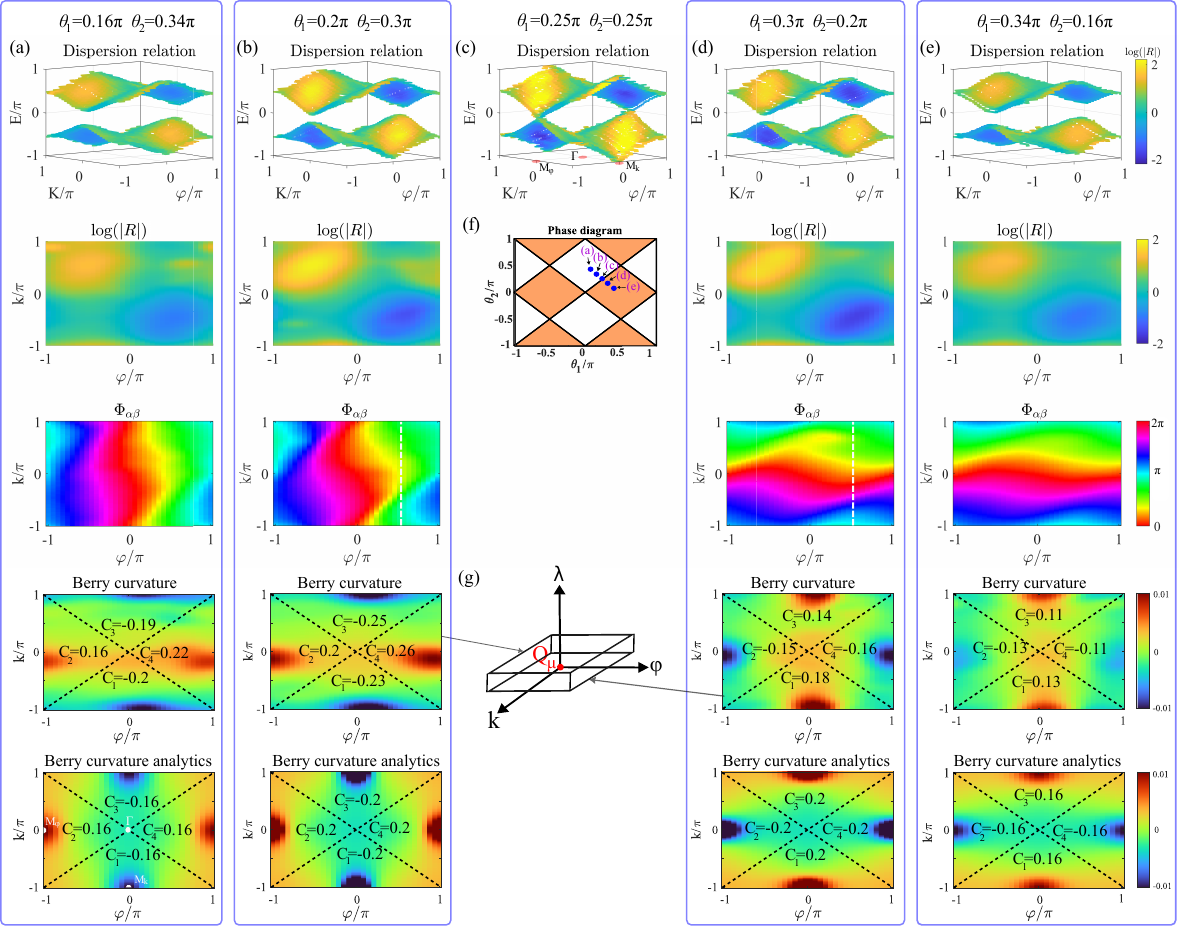}
\caption{\label{fig2} 
(a)-(b) Measured bands, ratio of amplitudes $\lvert R\rvert$, and phase difference $\Phi_{\alpha\beta}$ of the sublattice components of the upper band eigenvectors in the white phase in panel (f): (a) $\theta_1=0.16\pi$, $\theta_2=0.34\pi$; (b) $\theta_1=0.2\pi$, $\theta_2=0.3\pi$. The color code in the uppermost panels depicts the measured value of $\lvert R\rvert$ in $\log$ scale. The two lowermost panels display the measured and analytically calculated Berry curvature of the upper band. (c) Measured bands at the point where the $E=0$ and $E=\pi$ gaps close ($\theta_1=\theta_2=0.25\pi$). (d)-(e) Same as (a)-(b) for $\theta_1=0.3\pi$, $\theta_2=0.2\pi$ and $\theta_1=0.34\pi$, $\theta_2=0.16\pi$, respectively, in the orange phase. (f) Topological phase diagram of the system as a function of the splitting angles $\theta_1$ and $\theta_2$. Black lines indicate the simultaneous closure of the $0$- and $\pi$-gaps, which happens for $\theta_1 \pm \theta_2 = n \pi$, with $n\in\mathbb{Z}$. (g) Cuboid in parameter space for the computation of the topological charge.
}
\end{figure*}
%******************************************
%****************************************
%****************************************

%Experimentally, the eigenvalues and eigenvectors can be directly measured in the following way. %A square laser pulse of 1~ns in length is injected in the left ($\alpha$) ring. This pulse duration corresponds approximately to the size of a single site of the time multiplexed lattice and it populates all eigenstates. As it evolves, the initial excitation expands through both sublattices (see Fig.~\ref{fig1}(c),(d)).

To determine the eigenvalues and eigenvectors, the phase and amplitude of $\alpha_n^{m}$ and $\beta_n^{m}$ are measured by interfering the light extracted from the rings at each synthetic position and time step with a high coherence laser shifted in frequency by $\Omega=3$GHz from the injected laser pulse (see interference fringes in the inset of Fig.~\ref{fig:spatiotemporal}(a)).
A Fourier transform of the stroboscopic spatiotemporal dynamics of each ring at time steps corresponding to integer Floquet periods ($m=2,4,6,\cdots$),  is depicted in Fig.~\ref{fig:spatiotemporal}(c),(d). 
The panels directly show the Floquet-Bloch eigenvalues of the lattice (identical for the two sublattices) for a given value of $\varphi$.
Two bands separated by two gaps centered at $E=0$ and $E=\pi$ can be identified.
The complex Fourier transform contains information on the ratio of amplitudes $R$ and phase difference $\Phi_{\alpha\beta}$ between the two sublattice sites for each quasimomentum eigenvector of each band (see Appendix~\ref{app:data_processing} and Ref.~\cite{Lechevalier2021} for a detailed description of the eigenvector measurement technique). % can be directly extracted from the Fourier transform of the measured spatiotemporal evolution at time steps corresponding to integer Floquet periods ($m=1,3,5,\cdots$)~\cite{Lechevalier2021}.
This allows a complete experimental characterisation of the eigenvectors of the lattice for each band eigenvalue labelled by $k$ and $\varphi$: %Eq.~(\ref{Eq:step}):
\begin{align}\label{eigenmodes}
    \begin{bmatrix} 
           \alpha_{n}^{m}(k,\varphi) \\
           \beta_{n}^{m}(k,\varphi) \\
    \end{bmatrix}
  = \frac{1}{\sqrt{1+|R|^2}} \begin{bmatrix}  
           1 \\
           |R| e^{\Phi_{\alpha\beta}} \\
  \end{bmatrix}
  e^{i k n/2}  e^{i E(k,\varphi) m/2 }.
\end{align}

Depending on the splitting ratios $\theta_{1}$ and $\theta_{2}$, the system presents two distinct gapped phases. They are shown in Fig.~\ref{fig2}(f) in white and orange colours, and separated by lines at which both gaps close simultaneously. 
The upper panels in Fig.~\ref{fig2}(a)-(b) show the measured two-dimensional quasienergy bands at two different points in the white phase in panel (f).
The second and third panels in Fig.~\ref{fig2}(a)-(b) display the measured tomography of the eigenvectors of the upper band via the measured ratio of amplitudes $\lvert R\rvert$ and phase difference $\Phi_{\alpha\beta}$ between the two sublattices for each point in the Brillouin zone.
The values of $\lvert R\rvert$ present a dipole shape, with eigenvectors with a high weight in the $\beta$ sublattice at the upper left corner of the Brillouin zone, and a low weight at the lower right corner.
At $\theta_1=\theta_2=\pi/4$ both middle and upper gaps close simultaneously (see panel (c)), and open again in the orange phase of Fig.~\ref{fig2}(f), which measured eigenvectors are displayed in panels (d)-(e).
Interestingly, in the white region (panels (a)-(b)) the phase $\Phi_{\alpha\beta}$ winds along the $\varphi$ direction, while it winds along the $k$ direction in the orange phase corresponding to panels (d)-(e).
We will see below that this difference in winding of the sublattice phase in the two regions is directly connected to the existence of edge states for particular edge realizations.

%%%%%%%%%%%%%%%%%%%%%%%%%%%%%%%%%%%%%%%%%%%%%%%%%%%%%%
%%%%%%%%%%%%%%%%%%%%%%%%%%%%%%%%%%%%%%%%%%%%%%%%%%%%%%
\section{Topological charge associated to the phase transition\label{Sec:charge}}
%%%%%%%%%%%%%%%%%%%%%%%%%%%%%%%%%%%%%%%%%%%%%%%%%%%%%%
%%%%%%%%%%%%%%%%%%%%%%%%%%%%%%%%%%%%%%%%%%%%%%%%%%%%%%

Once the eigenvectors have been fully characterised via the amplitude ratio $\lvert R\rvert$ and the phase $\Phi_{\alpha\beta}$, it is straightforward to extract the Berry curvature at each point of the Brilluoin zone and the Chern number~\cite{blanco_de_paz_tutorial_2020}. 
The lower panels of Fig.~\ref{fig2} show the measured and analytically computed Berry curvature for the upper band in panels (a)-(b) and (d)-(e) (see Appendix ~\ref{app:data_processing} for a detailed description of the computation of the Berry curvature). 
%At the gap closing point, the bands are not isolated and the Berry curvature is not well defined.
The measured Chern number is approximately zero in all cases (values -0.01 in (a), 0.02 in (b), 0.01 in (d), 0.00 in (e)).
Nevertheless, a strong local Berry curvature is observed around the gap closing points $M_k$ and $M_\varphi$, shown in the lowermost panel of Fig.~\ref{fig2}. The local Berry curvature changes sign when going through the gap closing transition.

Despite the zero Chern number in the two phases, the local change of sign of the Berry curvature across the phase transition is a direct manifestation of the topological nature of this transition. This feature is captured by the topological charge $Q_{\mu}$ associated to the gap closing singularities of the quasi-energies ($\mu$ labels the considered gap).
For a given band, this charge can be determined from the outgoing Berry flux through a cuboid in parameter space ($k, \varphi, \lambda$) centred at the band touching singularity at the phase transition~\cite{unal_how_2019,Wintersperger2020}. This topological charge thus corresponds to a monopole Chern number, which is different from the band Chern number obtained from the Berry flux through the first Brillouin zone, which vanishes here.
The cuboid is represented in Fig.~\ref{fig2}(g) and spans a part of quasimomentum space ($k, \varphi$) with the third dimension $\lambda$ parametrizing the changes in $\theta_1$ and $\theta_2$ to go through the transition. 
The outgoing Berry flux in the upper and lower faces can be directly extracted from the integration of the Berry curvature around the high symmetry point at which the band touching point takes place. 
The flux through the lateral faces can be neglected if $\Delta \lambda$ is small enough, that is, if the Berry curvature in momentum space is measured at points sufficiently close to the phase transition. 
For the other band, the topological charge associated to the gap closing transition has an opposite sign.

To determine the topological charge at the $\mu=0$ gap, we evaluate the integrated Berry flux $B_f$ around the gap closing point $M_\varphi$ experimentally and analytically. 
To do so in practice, we divide the Brillouin zone in four regions indicated by dashed lines in the lower panels of Fig.~\ref{fig2} and integrate the measured Berry curvature over regions 2 and 4, which surround the $M_\varphi$ point. 
The white region side corresponds to the upper face of the cuboid. Close to the phase transition (Fig.~\ref{fig2}(b)), the integrated outgoing flux is $B_f=(C_2+C_4)=0.46$. 
At the orange side depicted in Fig.~\ref{fig2}(d), it is $(-1)\times(-0.31)$.
The prefactor $-1$ comes from the fact that we evaluate the flux exiting the cuboid through the lower face (Fig.~\ref{fig2}(g)).
The total measured flux is $0.77$, close to the value of 0.80 computed from the analytic eigenvectors and represented in the lowermost panels of Fig.~\ref{fig2}(b),(d). 
The expected value of the topological charge is 1.
Analytic evaluation of the Berry curvature on the lateral sides of the cuboid shows that the Berry flux through these faces is zero (see Fig.~\ref{fig15} in the appendices).
Therefore, the difference with the measured values stems from the spread of the Berry curvature associated to the $M_\varphi$ point at finite gap sizes. 
By studying the Berry flux at points closer to the phase transition, analytic calculations show that the value of $Q_{0}$ extracted from the local Berry flux approaches 1. 
To study a situation with a smaller gap experimentally we are limited by the resolution of the quasienergy bands, which is given by the number of accessible round trips (84 in our case).

The same treatment can be done for the Berry flux around the closing point for the $\mu=\pi$ gap. To preserve the correct sign convention we need to evaluate the Berry flux on the lower band around the $M_k$ point. We obtain the experimental value $B_f^{+\lambda}-B_f^{-\lambda}= -0.80$ (see Appendix \ref{app:data_processing}). Though the measured Berry fluxes depart from the value of 1 expected for the topological charge at the transition, they are sufficiently different from zero to allow an unambiguous determination of the topological nature of the phase transition.

A similar analysis was recently done in an ultracold atoms experiment implementing a two-bands, two-dimensional lattice subject to periodic shaking~\cite{Wintersperger2020}. In that case, the studied topological phases involved phase transitions with the closure of a single gap at a time, and a change of the number of edge states in that specific gap through the transition. To determine the topological charge related to the gap closing singularity, the Berry flux was computed across the whole Brillouin zone, that is, via the change of Chern number in the bands across the transition. In our case, both gaps close and open simultaneously and the Chern number of the bands is zero in all topological regions. Our work shows that a local analysis of the Berry flux around each gap closing point allows an accurate characterisation of the topological charge.

%****************************************
%****************************************
%****************************************
\begin{figure}[t!]
\includegraphics[width=\columnwidth]{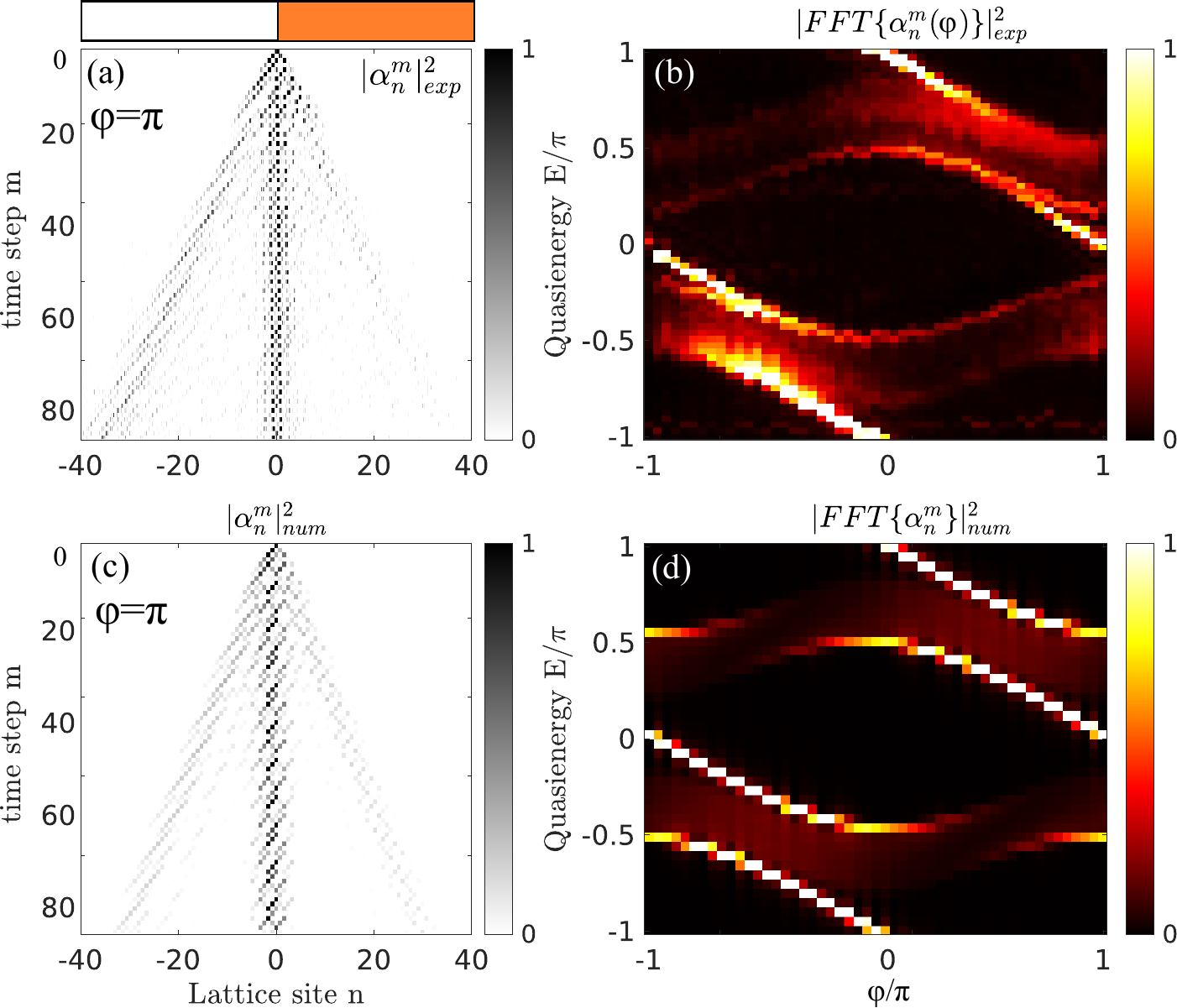}
\caption{\label{fig4} 
(a) Measured step walk when exciting a single site at the interface between a lattice in the white region ($\theta_1=0.2\pi,\theta_2=0.3\pi $) and a lattice in the orange region ($\theta_1=0.3\pi,\theta_2=0.2\pi $) for $\varphi=\pi$. (b) Measured band structure as a function of $\varphi$ showing an interface state traversing both gaps. (c) and (d) Numerical simulation of Eq.~(\ref{Eq:step}) in the conditions of the experiment.
}
\end{figure}
%******************************************
%****************************************
%****************************************

The direct consequence of the non-trivial topological charge at the gap closing singularities is the existence of interface states when lattices in the two different topological regions are pasted together. We implement a lattice with two different spatial regions by engineering the splitting ratios in the proper way on either side of the boundary~\cite{Adiyatullin2023}. The top inset of Fig.~\ref{fig4} shows a sketch of the employed lattice: the left part belongs to the white phase with $\theta_1=0.2\pi$ and $\theta_2=0.3\pi$, while the right part belongs to the orange phase with $\theta_1=0.3\pi$ and $\theta_2=0.2\pi$. By injecting an initial pulse exactly at an interface site, we observe that part of the light remains localised at the interface (Fig.~\ref{fig4}(a)). Figure~\ref{fig4}(b) shows the measured bands when scanning the values of $\varphi$ from $-\pi$ to $\pi$ and keeping the injection at the interface. Each gap is traversed by a single band of interface states, with the same group velocity in both cases, a situation that consistently corresponds to an anomalous topological phase.
%This matches perfectly the expectation from the change of windings across the phase transition for both gaps, which in both cases is equal to the measured topological charge close to 1, as discussed above.
%%%%%%%%%%%%%%%%%%%%%%%%%%%%%%
%%%%%%%%%%%%%%%%%%%%%%%%%%%%%%
\begin{figure*}[t!]
    \includegraphics[width=\textwidth]{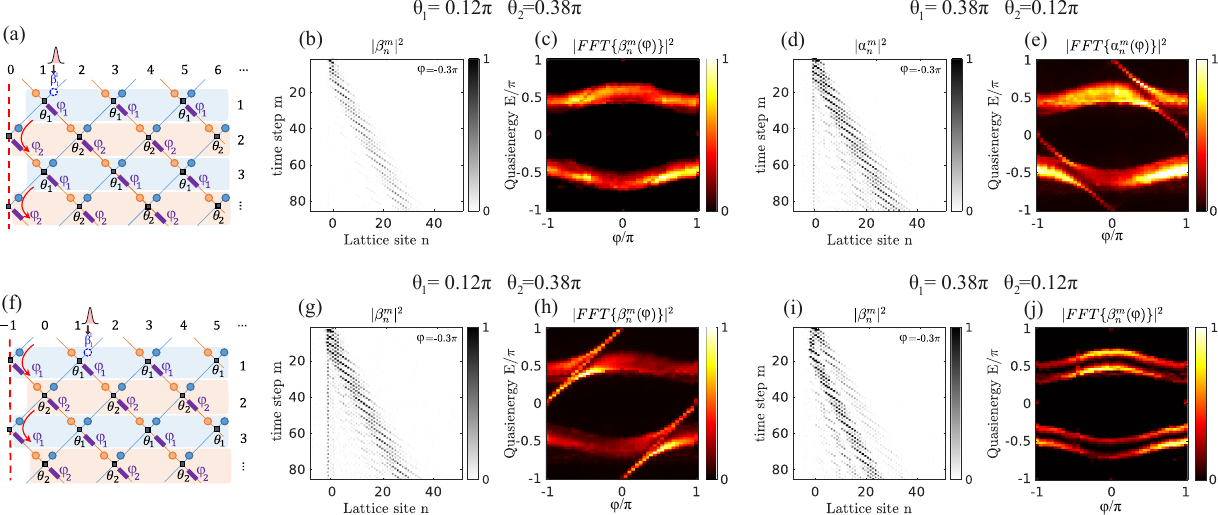}
    \caption{(a) Schematics of the quantum walk with a fully reflecting edge along the position coordinate at site $n=0$ and periodic boundary conditions along the synthetic dimension, $\varphi$. The two-step protocol separates the stroboscopic evolution into odd and even time-steps (blue and red time slices, respectively). Their corresponding Floquet operators have identical quasienergies, but they describe different time-frames, or equivalently, different boundary conditions. 
    (b), (d) Measured spatiotemporal dynamics in the $\beta$ (b) and $\alpha$ (d) rings when injecting an initial pulse at the $\beta$ ring at site $n=1$ in the white phase (b) and orange phase (d) of Fig.~\ref{fig2}(f), for $\varphi=-0.3\pi$. Only in the second case an edge state band traverses both gaps, as displayed in the measured dispersions in (c) and (e).
    To compose these panels, we first Fourier transform the measured spatioteporal dynamics selecting only the even time steps (stroboscopic analysis) for each value of $\varphi$. We then stack the resulting bands for different values of $\varphi$.
    The edge state is visible in the spatiotemporal diagram in (d) as a localised emission at the boundary with vacuum.
	(f)-(j) Same as the upper row (a)-(e) when adding to the lattice an extra site $n=-1$ at the edge with vacuum. The injection site in all experimental images is still $n=1$. The addition of a site to the lattice exchanges the presence/absence of chiral edge states with respect to the case depicted in the upper row.}
    \label{fig:edge}
\end{figure*}
%%%%%%%%%%%%%%%%%%%%%%%%%%%%%%
%%%%%%%%%%%%%%%%%%%%%%%%%%%%%%
%%%%%%%%%%%%%%%%%%%%%%%%%
\section{Bulk Topological characterization}
%%%%%%%%%%%%%%%%%%%%%%%%%
The previous section experimentally demonstrates the presence of anomalous topology in the implemented lattice, but we have not yet characterised the topology of the different phases. 
In contrast with Floquet systems whose dynamics are governed by continuous time-modulations of a Hamiltonian~\cite{Rudner2013}, it has been recently discussed that the presence of edge states in discrete step walks -- especially in the anomalous regime -- is not fully characterized by a bulk invariant~\cite{Delplace_Fruchart_Tauber,delplace_topological_2020,Bessho2022}.
In other words, the existence of edge states depends on the boundary conditions. This is somehow very similar to static chiral symmetric systems in one dimension such as the the Su-Schriefer-Heeger (SSH) model, except that the SSH model is 1D, while the boundary conditions dependence for the edge states in discrete steps walks was noticed in higher dimensions, and in particular for chiral edge states in 2D. 

Different approaches have been followed to tackle the issue of boundary dependent edge states to still have a topological meaning. 
One strategy consists in noticing that the boundary conditions are in correspondence with different choices of unit cells that define different bulk evolution operators. 
Although those operators yield different bulk invariants, their difference is still meaningful to characterise interface states \cite{Delplace_Fruchart_Tauber}. 
This situation corresponds to the description in terms of topological charge at the gap closing points we followed in Sec.~\ref{Sec:charge}.
A second option is to identify the winding of the discrete step walk sketched in Fig.~\ref{fig1}(b) if interpreted as a finite oriented graph in two dimensions, in which one of the dimensions is the position and the second dimension the step time\cite{delplace_topological_2020}.
This graphical approach is equivalent to computing a quantized flow of the boundary modes through an arbitrary line transverse to the edge in finite size geometry, and it is particularly suitable in 2D but does not seem straightforward to apply to our 1D case plus time step.
A third approach, dubbed extrinsic topology, was recently proposed to  handle the boundary condition dependence by separating the bulk evolution from the edge evolution, and considering the topological properties of both bulk and edge operators \cite{Bessho2022}. In this approach, ''extrinsic'' refers to the boundary of the system, which is not fixed, in contrast with the  ''intrinsic'' topology that denotes the bulk. This method seems general enough but the edge operator may be tricky to identify in practice.

In the present work we follow a yet different strategy by exploiting the fact that the choice of boundary conditions is not only related to a choice of bulk unit cell, as it is in certain static Hamiltonian systems (e.g. the SSH Hamiltonian), but also to a time frame, which is specific to discrete step walks. This property can be related to what was introduced as time-glide symmetry \cite{Mochizuki2020}.
In this manner, fixing a time frame, that is, the origin of time over which we have an experimental control, sets the bulk evolution operator in a consistent way with the boundary conditions we may implement.

In the following, we will develop this approach and find a suitable topological invariant that takes into account the choice of unit cell/time frame associated to the specific boundary conditions. For completeness,  in Appendix~\ref{appendix:AppendixSSH} we demonstrate how the extrinsic topology approach yields the same phase diagram, and in Appendix~\ref{appendix:Glide} we show that the invariant derived for systems with time-glide symmetry coincides with the one we propose in this work.

%%%%%%%%%%%%%%%%%%%%%%%%%
\subsection{Bulk Topology}
The bulk topology can be determined from the Floquet operator under Periodic Boundary Conditions (PBC). 
The Fourier transform of Eq.~(\eqref{eq:EOM1}) for the 2-step protocol results in the following Floquet operator:
%%%%
\begin{align}
U_{\text{QW}}(\mathbf{q})&=v_{0}(\mathbf{q})\sigma_{0}+i\mathbf{v}(\mathbf{q})\cdot\boldsymbol{\sigma},\label{eq:UQW1}
\end{align}
with the two-dimensional vector $\mathbf{q}=(k,\varphi)$. The components of the vector $\mathbf{v}(\mathbf{q})$ can be expressed in the following form (details in the Appendix\ref{appendix:Invariant}):
%%%
\begin{align}
%v_{0}&=J_{1}\cos\left(k\right)-m_{1}\cos\left(\varphi\right),\\
v_{x}&=J_{2}\cos\left(k\right)+m_{2}\cos\left(\varphi\right),\\
v_{y}&=J_{2}\sin\left(k\right)+m_{2}\sin\left(\varphi\right),\\
v_{z}&=-J_{1}\sin\left(k\right)+m_{1}\sin\left(\varphi\right)
\end{align}
%%%
with the different parameters being functions of the angles $\theta_{1,2}$ of the variable beamsplitter:
\begin{align}
    J_{1}=& \cos\left(\theta_{1}\right)\cos\left(\theta_{2}\right), J_{2}=\sin\left(\theta_{1}\right)\cos\left(\theta_{2}\right), \nonumber\\
    m_{1}=& \sin\left(\theta_{1}\right)\sin\left(\theta_{2}\right),\ m_{2}=\cos\left(\theta_{1}\right)\sin\left(\theta_{2}\right).
\end{align}
Actually, those four parameters are not independent due to the unitarity of the time-evolution operator that imposes the constrains $J_{1}m_{1}=J_{2}m_{2}$ and $J_{1}^{2}+J_{2}^{2}+m_{1}^{2}+m_{2}^{2}=1$.

Importantly, to write this Floquet operator, we have assumed that during the first and second time-steps the beamsplitter is set to $\theta_{1,2}$, respectively.
This allows us to identify the 2-step stroboscopic evolution generated by $U_\text{QW}(\mathbf{q})$ with the blue time-slices in the schematic of Fig.~\ref{fig:edge}(a).
However, due to time-periodicity, another perfectly valid choice would have been to consider a shifted time-frame in which the beamsplitter is set to $\theta_{2,1}$ during the first and second time-steps, respectively.
We will refer to this alternative choice as $\tilde{U}_{\text{QW}}(\mathbf{q})$, which corresponds to the red time-slices of the evolution in Fig.~\ref{fig:edge}(a).
Their matrix representation differs, but they are connected by the swap of the angles $\theta_1 \leftrightarrow \theta_2$, plus an additional sign change in the quasimomentum $\varphi\to -\varphi$.
Their equivalence becomes more clear when one checks that their quasienergy spectrum is identical, and therefore, $\tilde{U}_\text{QW}(\mathbf{q})$ and $U_\text{QW}(\mathbf{q})$ have identical Floquet bands.
This symmetry between frames with an additional reflection along the synthetic dimension $\varphi$ can be related with a recently proposed symmetry relevant to the topology of quantum walks, known as time-glide symmetry~\cite{Mochizuki2020}, which we will not discuss further here.

To find the corresponding bulk topological invariant, one can first check that $U_\text{QW}(\mathbf{q})$ is in the D symmetry class with particle-hole symmetry
\begin{equation}
    \mathcal{C}U_{\text{QW}}(\mathbf{q})\mathcal{C}^{-1}=U_{\text{QW}}(-\mathbf{q}),
\end{equation}
implemented by the anti-unitary operator $\mathcal{C}=\sigma_z K$, being $K$ the complex conjugation operation. This implies that the bulk invariant corresponds to the Chern number~\cite{Ryu_2010}.
The numerical calculation of the Berry curvature distribution perfectly agrees with the one measured in the experiment (see Fig.~\ref{fig2}), and vanishes when summed over the whole Brillouin zone, for all values of $\theta_{1,2}$. Hence, the Chern number of the Floquet bands is zero, and this indicates that only trivial or anomalous phases are present in our system.
In addition, we have experimentally demonstrated above that the two phases separated by a gap closure are topologically distinct, because chiral edge states develop at their interface (see Fig.~\ref{fig4}), together with a change of sign of the local Berry curvature captured in the topological charges $Q_\mu$.
Hence, one can conclude that, although the Chern number vanishes, the phase diagram must contain at least one anomalous Floquet phase.

\subsection{Reference frames and open boundary conditions}

To identify each phase independently, we perform an additional experimental test where we implement open boundary conditions (OBC) with the vacuum along the spatial coordinate and PBC along the synthetic one.
OBC are experimentally implemented by dynamically changing the variable beamsplitter at a single site, and setting its value to full reflectance (see Fig.~\ref{fig:edge}(a) for a schematic).
Figure~\ref{fig:edge}(b)-(e) shows the presence of robust chiral edge states in one phase (orange panels in Fig.~\ref{fig2}(f)) and the absence of edge states in the other (white panels in Fig.~\ref{fig2}(f)). Numerical simulations of Eq.~(\ref{Eq:step}) confirmed the experimental observations (see Fig.~\ref{fig:edge_sim} in the Appendix).
This behaviour seems to indicate that the white panels of the phase diagram can be identified with the trivial phase, while the orange panels with the anomalous phase.

However, this appears to be in contradiction with the time translation invariance between $U_\text{QW}(\mathbf{q})$ and $\tilde{U}_\text{QW}(\mathbf{q})$ previously discussed.
The reason is that the two phases considered in Fig.~\ref{fig:edge}(b)-(c) and (d)-(e) are related by the swap $\theta_1\leftrightarrow\theta_2$.
Hence, they are described by equivalent Floquet operators whose only difference is a shift in the reference frame by one step of the protocol and this should not change the bulk topology.
That is, one would expect the two phases to be identical because they are related by a symmetry of the system and, hence, no edge states should exist at their interface.
This is consistent with the fact that the Chern number remains zero at both sides of the quasienergies gap closure.
We now resolve this contradiction by demonstrating the crucial role played by the boundaries.

To understand the role of the boundaries in the presence of edge states, notice that the choice of a time-frame in Fig.~\ref{fig:edge}(a) can be linked to a choice of boundary when OBC are considered.
For example, if one chooses $U_{\text{QW}}(\mathbf{q})$ to characterize the stroboscopic evolution (blue time-slices in Fig.~\ref{fig:edge}(a)), this defines a bipartite unit cell $(\alpha_n,\beta_n)$. There is nothing special about this unit cell, as far as PBC are considered along the spatial dimension. 
However, once the fully reflecting edge is fixed at $n=0$, the evolution can be interpreted as the stroboscopic evolution of a chain of dimers with an $\alpha$-site termination.
Instead, if one chooses the Floquet operator $\tilde{U}_{\text{QW}}\left(\mathbf{q}\right)$ to characterize the stroboscopic evolution (red time-slices in Fig.~\ref{fig:edge}(a)), for a boundary at the same site, this corresponds to a dimer chain with a $\beta$-site termination.
This implies that for OBC, a change in the boundary termination is equivalent to a shift in the reference frame of the Floquet operator.
This is shown schematically in Fig.~\ref{fig:edge}(a), where the shift in one time-step (change of color in the time-slice) is equivalent to a change of boundary.

This relation between different edge terminations for OBC and Floquet operators with PBC in different frames of reference provides the starting point to obtain the topological invariant from the Floquet operator with PBC incorporating the role of boundaries.

As a confirmation that our interpretation is correct regarding the importance of boundaries, and that there is nothing special about the choice of $U_{\text{QW}}(\mathbf{q})$ or $\tilde{U}_{\text{QW}}(\mathbf{q})$ (i.e., they have identical bulk topology), we show in Fig.~\ref{fig:edge}(f)-(j) that a change in the boundary to a fully reflecting edge at site $n=-1$ for the same values of $\theta_{j}$, reverses the phase diagram shown in Fig.~\ref{fig2}(f).
Hence, as we can find chiral edge states for any value of $\theta_j$, this confirms that the system can actually be topological for all values of $\theta_j$, and that the appearance of the chiral edge states is linked to the particular boundaries fixed in the experimental realization.
Also, notice that the direction of the group velocity associated to the edge state in Fig.~\ref{fig:edge}(h) is reversed with respect to the case with a boundary at site $n=0$ in Fig.~\ref{fig:edge}(e). The reason is that, as we discussed earlier, the time translation relating the two frames is not only implemented by the swap $\theta_1 \leftrightarrow \theta_2$, but it also involves the reflection of the quasimomentum $\varphi \to -\varphi$. Therefore, as changing the boundary site must be equivalent to a reference frame shift, the phase diagram not only reverses the phases, but also the chiral edge states change their direction of propagation.

\subsection{Topological invariant}

\begin{figure}[t]
    \includegraphics[width=0.65\columnwidth]{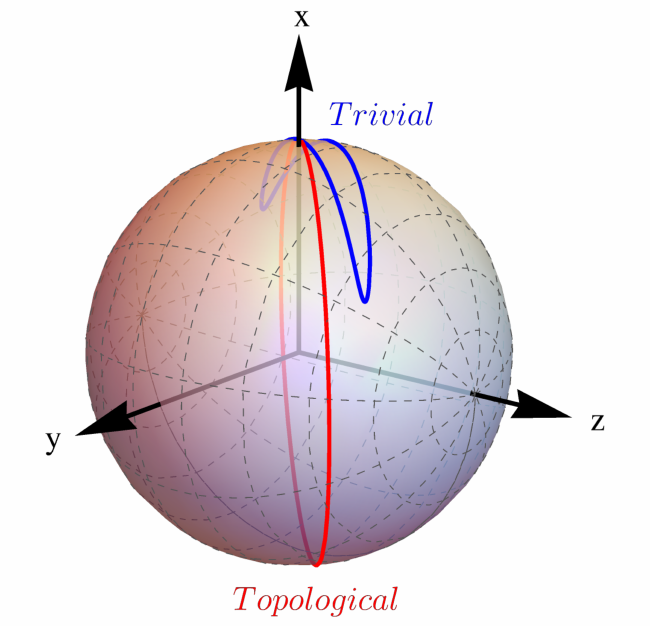}
    \caption{Topologically nonequivalent trajectories on the Bloch sphere as $k$ is varied over the first Brillouin zone.
    Particle-hole symmetry requires that at high symmetry points $\varphi=m\pi$, for all $m\in\mathbb{Z}$, the Bloch vector aligns with the $x$-axis if $k = \{ 0,\pi \}$. This defines two nonequivalent loops that either cross twice the same pole (trivial, in blue) or once each different pole (topological, in red). These loops cannot be continuously transformed into one-another. For visualization purposes we chose $\varphi=0.99\pi$, which is why loops in solid lines are not exactly confined to a plane when chiral symmetry is present.}
    \label{fig:trajectories}
\end{figure}

For the purpose of defining the topological invariant and predicting the existence of chiral edge states for a particular boundary, we now focus on the time frame that defines the evolution operators $U_{\text{QW}}(\mathbf{q})$ and study the conditions to have edge states when OBC are placed along the spatial coordinate. %That is, we are interested in non-trivial loops in $q$-space (Confusing, we need to discuss this sentence).
This evolution operator describes the stroboscopic evolution depicted by blue time-slices in Fig.~\ref{fig:edge}(a).
Importantly for our purpose, particle-hole symmetry imposes a crucial restriction on the trajectories of the Bloch vector $\mathbf{v}(\mathbf{q})$ when $k$ and $\varphi$ are varied. It requires that at high-symmetry points $\mathbf{q}=\left(\pi n, \pi m \right)$ with $n,m=\left\{ 0, \pm1 \right\}$, the Bloch vector aligns with the $x$-axis, $\mathbf{v}=(v_x,0,0)$.
This allows us to define two topologically nonequivalent trajectories~\cite{Sato2017}, schematically shown in Fig.~\ref{fig:trajectories}: blue trajectories passing through a single pole of the Bloch sphere only, and red trajectories passing through both poles.
%%%%%%%%%%%%%%%%%%%%%%%%%%%%%%
%%%%%%%%%%%%%%%%%%%%%%%%%%%%%%
%%%%%%%%%%%%%%%%%%%%%%%%%%%%%%
%%%%%%%%%%%%%%%%%%%%%%%%%%%%%%

To define the topological invariant we focus in loops in $k$-space. 
We first consider the particular case of the high symmetry point $\varphi=m \pi$, with $m\in\mathbb{Z}$.
We will show that a non-zero winding number exists in this case, and then demonstrate that our argument holds for arbitrary $\varphi$, allowing to discriminate between the two types of topologically nonequivalent loops that perfectly predict the phase diagram of Fig.~\ref{fig2}(f).

At the high symmetry points the system has an additional symmetry because $U_{\text{QW}}(\mathbf{q})$ becomes chiral (the explicit form of the chiral operator can be found in Appendix~\ref{appendix:Invariant}):
%%%%%%%%%%%%%%%%%%%%%%%%%%%%%%
\begin{equation}
    \Gamma U_{\text{QW}}(k,n\pi) \Gamma^{-1} = U_{\text{QW}}(k,n\pi)^{\dagger}.
\end{equation}
%%%%%%%%%%%%%%%%%%%%%%%%%%%%%%
Then, one can calculate the winding number that distinguishes between trajectories of the Bloch vector $\mathbf{v}$  that encircle the origin from those that do not, while $k$ spans the Brillouin zone~\cite{Asboth2013}. This invariant is given by counting algebraically the number of intersections of this trajectory with an arbitrary vector $\mathbf{v}_0$ pointing to the equator from the center of the Bloch sphere, as
\begin{align}
    \nu =\sum_{\mathbf{v}(k_i)=\mathbf{v}_0} \text{sgn} [(\mathbf{v}\times \partial_k\mathbf{v}) \cdot \hat{\mathbf{n}}]
\end{align}
where $k_i$ denote the momenta at which an intersection $i$ occurs, and $\hat{\mathbf{n}}$ is a unit vector normal to the equatorial plane~\cite{Sticlet2012,Quandt2021}. Furthermore, as we know that at $k=\left\{0,\pi\right\}$ the Bloch vector must align with the $x$-axis due to particle-hole symmetry, we can choose $\mathbf{v}_0$ as pointing along the $x$ direction, and this yields the expression  (see Appendix~\ref{appendix:Invariant} for a detailed derivation):
%%%%%%%%%%%%%%%%%%%%%%%%%%%%%%
\begin{equation}
     \nu = \frac{1}{2}\left[ 1-\text{sgn}\left(J_{2}+m_{2}\right)\text{sgn}\left(m_{2}-J_{2}\right)\right].\label{eq:winding2}
\end{equation}
%%%%%%%%%%%%%%%%%%%%%%%%%%%%%%
Here each sign function determines if the Bloch vector aligns with the north or the south pole at $k= \{ 0,\pi \}$. Hence, it shows that trajectories that cross the two poles wind around the origin, and those that cross the same pole twice, do not wind.

To generalize our argument to arbitrary $\varphi$ we just need to notice that varying $\varphi\in\left[-\pi,\pi\right]$ does not close the quasienergies gap. Therefore, the loops are smoothly deformed out-of-plane (see Fig.~\ref{fig:CompositeTrajectories} in Appendix~\ref{appendix:Invariant}), and although the system is not chiral anymore, the two types of trajectories can still be topologically identified due to particle-hole symmetry.

%%%%%%%%%%%%%%%%%%%%%%%%%%%%%%
%%%%%%%%%%%%%%%%%%%%%%%%%%%%%%
\begin{figure}[t]
    \includegraphics[width=\columnwidth]{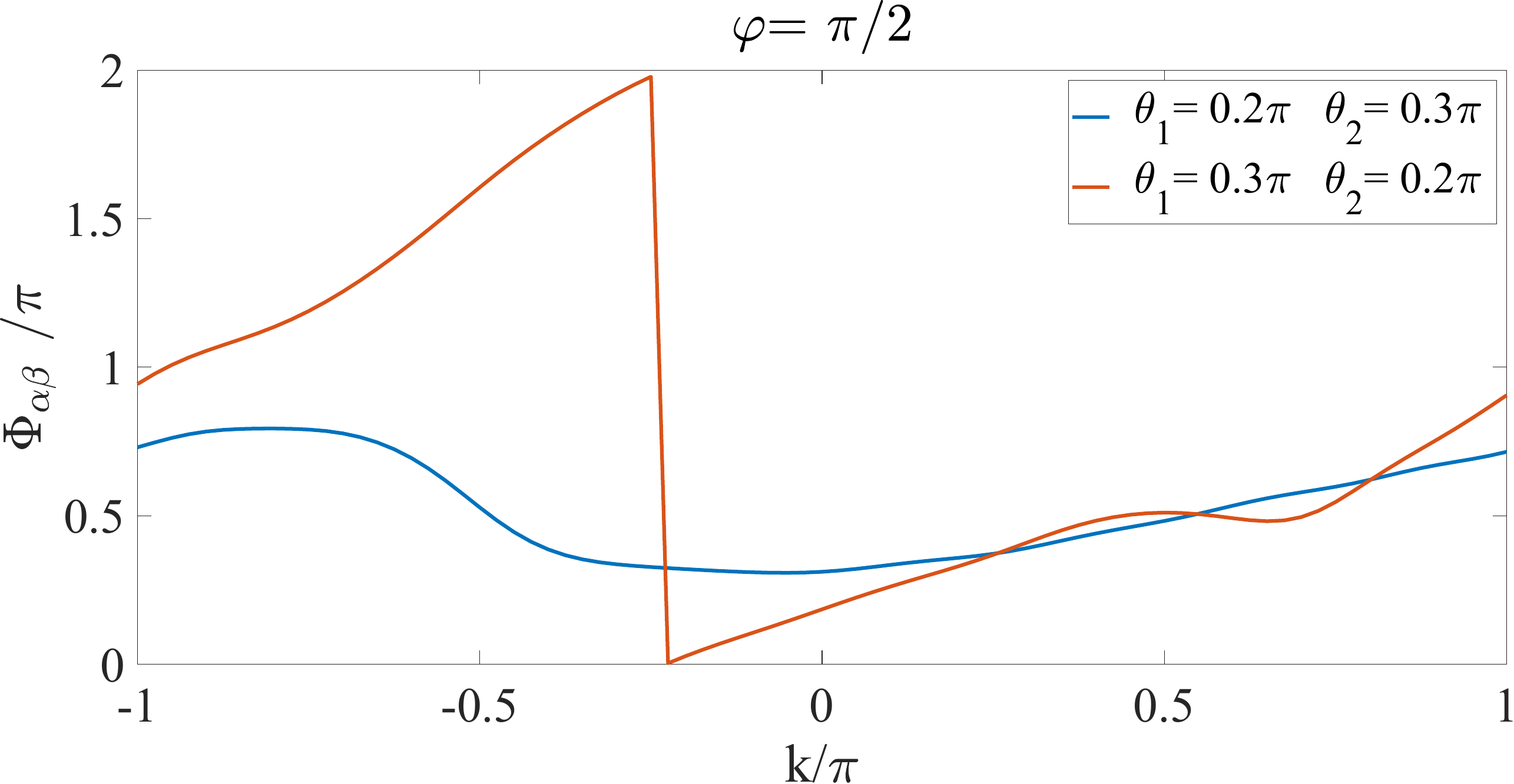}
    \caption{Measured value of the sublattice phase difference $\Phi_{\alpha\beta}$ for $\varphi = 0.5\pi$ extracted from panels (b) and (d) of Fig.~\ref{fig2} (dashed lines). The winding of $\Phi_{\alpha\beta}$ is clearly observed in the case with topological edge states, corresponding to the orange regions in Fig.~\ref{fig2}(f).}
    \label{fig:phasewinding}
\end{figure}
%%%%%%%%%%%%%%%%%%%%%%%%%%%%%%
%%%%%%%%%%%%%%%%%%%%%%%%%%%%%%

The invariant defined in Eq.~(\eqref{eq:winding2}) predicts the existence of edge states and the phase diagram of Fig.~\ref{fig2}(f). 
More importantly, it also allows extracting the value of the invariant directly from the experimental measurements.
The reason is that the vector $\mathbf{v}(\mathbf{q})$ characterizes the time-evolution operator in Eq.~(\eqref{eq:UQW1}), and its trajectory controls the relative phase between components of the eigenstates.
Figure~\ref{fig2} displays the measured sublattice phase difference of the eigenvectors, $\Phi_{\alpha\beta}$.
Hence, in the phase with edge states, the winding number must be $1$ for trajectories along $k$. This is precisely what is reported in Fig.~\ref{fig2}(d)-(e) for fixed $\varphi$.
In contrast, when $\theta_{1,2}$ swap their values, as in Fig.~\ref{fig2}(a)-(b), the system has identical quasienergies but no edge states. This is because the texture $\Phi_{\alpha\beta}$ of phase difference has rotated by $90$ degrees in parameter space $(k,\varphi)$, making the winding along $k$ trivial.
This is explicitly shown in Fig.~\ref{fig:phasewinding}, where we display $\Phi_{\alpha \beta}$ measured in the two phases. 
%between the phase with edge states and the one without edge states is extracted from the experiment.
%%%%%%%%%%%%%%%%%%%%%%%%%%%%%%%%%%%%%%%%%%%%%%%%%%%%%%%%%
%%%%%%%%%%%%%%%%%%%%%%%%%%%%%%%%%%%%%%%%%%%%%%%%%%%%%%%%%
\subsection{Connections with the topology of a chain of dimers}
As we mentioned above, the topology of the system seems to share some similarities with that of a chain of dimers (the SSH model)~\cite{Delplace2011,perez-gonzalez_interplay_2019} in the sense that the existence of edge states is linked to the type of considered boundary~\cite{Bernevig2013}.
Discussing this apparent similarity will allow us to place our results within the framework of extrinsic topology and explicitly show how edge unitaries affect the global topology of the system.

The unit cell choice of A-B or B-A atoms in a chain of static dimers is similar to the choice of the order of time steps in our setup. 
In both cases, this choice only fixes the spatial or temporal reference frame, respectively, which is unimportant for the physics of the bulk system. 
When OBC are considered the two cases become different. In a chain of static dimers, a non-vanishing winding number implies that an unpaired state is present at the edge of a chain. 
The winding number is now unambiguously defined for the particular choice of unit cell corresponding to the dimer at the edge. 
The existence of the edge state is then explained by the bulk-boundary correspondence of chiral symmetric Hermitian Hamiltonians in dimension one~\cite{Delplace2011,Jezequel22}, and the value of the invariant is linked to the choice of termination (i.e, the unit cell).

However, in the case treated in the present work, there is an important difference due to the Floquet dynamics.
The discrete nature of the time periodicity in our lattice enriches the topology with the possibility of the winding of the quasienergy spectrum. 
Mathematically, this is related to the fact that in Floquet physics, one must deal with a unitary operator -- the time-evolution operator -- and not a Hermitian one.
Hence, a non-vanishing winding number, $\nu\neq 0$, indicates the presence of a topologically protected unitary operator $U_\text{edge}(\varphi)$ that acts on the dimers at the edges for a particular edge termination, and this edge unitary has a non-trivial topology characterized by a winding number $\nu_{edge}$.
%That is how the edge topology contributes to the global topology of the system and how the extrinsic topology framework naturally emerges in this setup.
%characterized by:
%\begin{equation}
%    w_1[U_\text{edge}(\varphi)] = \int_0^{2\pi}\frac{d\varphi}{2\pi}\text{tr}[U_\text{edge}(\varphi)^{-1} i \partial_\varphi U_\text{edge}(\varphi)],
%\end{equation}
%which also contributes to the global topology of the system.
This additional contribution to the bulk or intrinsic topology by edge unitaries is what is called extrinsic topology~\cite{Bessho2021}.
In our case, the unitaries at the edge are effectively determined by the choice of time step (even or odd, or equivalently $\theta_1$ and $\theta_2$) at which the spatial unit cell is cut by the edge.
In Fig.~\ref{fig:edge}(a), the edge unit cell is cut at even steps, implicitly imposing the blue times reference frame (odd steps) for the determination of the edge unitaries. The opposite reference frame is imposed in Fig.~\ref{fig:edge}(f).
In Appendix~\ref{appendix:AppendixSSH} we explicitly find the edge unitaries for these cases and provide the details to characterize their appearance.
The change of unitaries and, therefore, the presence of edge states, can also be controlled by explicitly engineering the unitary operators themselves without changing the position of the edge.
This situation is not possible in static Hermitian systems, and shares features with non-Hermitian Hamiltonians, as discussed in Ref.~\cite{Bessho2021}.

Notice that extrinsic topology is a hallmark of time-discrete evolution, but plays a similar role as the micromotion in continuously driven Floquet Hamiltonians.
That is, extrinsic topology produces chiral edge modes simultaneously in both gaps, even when the bulk bands are topologically trivial, which is one of the signatures of anomalous topology when micromotion becomes relevant.
However, in extrinsic topology the control of these chiral edge modes depends on the unitaries at the edge sites. This situation is very different to the Floquet topology expected from continuously driven systems, in which anomalous phases are described by a purely bulk invariant~\cite{Rudner2013}.

%%%%%%%%%%%%%%%%%%%%%%%%%%%%%%
\section{Conclusions}
%%%%%%%%%%%%%%%%%%%%%%%%%%%%%%
In this work we have addressed a complete theoretical and experimental analysis of anomalous topological phases in a synthetic photonic lattice following a step-walk evolution.
Our system illustrates some fundamental differences between static and driven systems, but also between driven Hamiltonians and general step-walk evolutions.
As such, it also confirms the existence of topological phases with time-glide symmetry~\cite{Mochizuki2020}.

In particular, we have shown that systems following a discrete and periodic time-evolution display topological features beyond those of continuously driven Floquet Hamiltonians.
These features stem from the absence of an intrinsic time-coordinate in discrete step walks and enlarge the possible topological phases~\cite{Mochizuki2020}: anomalous phases and the emergence of chiral edge states are controlled by both the bulk topology and the specific edge termination. The existence of such "extrinsic" topology is demonstrated in our experiments.
To predict the presence of chiral edge states in our system we use a novel invariant tailored for two-dimensional step-walks in the D symmetry class. 
This invariant takes into account the crucial role of boundaries in extrinsic topology by showing that different time-frames become nonequivalent in the presence of OBC.
It allows identifying the different anomalous phases in the phase diagram without any reference to the micromotion and demonstrates that the presence of edge states is completely dominated by extrinsic topology.

From an experimental perspective, our results demonstrate that a complete topological characterization of anomalous phases in a synthetic photonic lattice is possible. We have determined the bulk/intrinsic topology of the synthetic lattice from the measurement of the eigenstates of the system, which are obtained from the Fourier transform of the spatial trajectories. 
From them, we extract the Berry flux distribution for each phase and the Chern number.
Additionally, we have shown that it is also possible to measure the topological charge associated to an arbitrary quasienergy band crossing at the $0$- or $\pi$-gap. This is helpful to study both, standard and anomalous topological phases.
For the study of edge/extrinsic topology we have shown the possibility to simultaneously detect the edge states and measure the winding number that relates the existence of edge states and the particular boundary realizations.

Our experimental platform paves the way for the investigation of genuine step walk topology, including the potential observation of extrinsic topological phases when arbitrary unitary operators are introduced at the edge sites~\cite{Bessho2022}.
These unitary operators can be designed to have local windings (i.e., at edge sites) by simply applying an appropriate voltage sequence to the phase modulator already present in one of the rings. 
This is not only interesting from a fundamental point of view, but it can also open new avenues to produce topological phases on demand, in situations in which the control of the bulk topology can be challenging. It could be of potential use to switch on and off topological channels via local actions at the edges of a two-dimensional metamaterial.

\begin{acknowledgments}
We thank Nathan Goldman for insightful discussions and Rajesh Asapanna for careful reading of the manuscript. This work was supported by the European Research Council grant EmergenTopo (865151), the French government through the Programme Investissement d'Avenir (I-SITE ULNE /ANR-16-IDEX-0004 ULNE) managed by the Agence Nationale de la Recherche, the Labex CEMPI (ANR-11-LABX-0007) and the region Hauts-de-France.
A.G.L. acknowledges support from the Proyecto Sinérgico CAM 2020 Y2020/TCS-6545 (NanoQuCo-CM) and the CSIC Research Platform on Quantum Technologies PTI-001.
\end{acknowledgments}
%****************************************

\bibliography{biblio}% Produces the bibliography via BibTeX.

\begin{thebibliography}{57}%
\makeatletter
\providecommand \@ifxundefined [1]{%
 \@ifx{#1\undefined}
}%
\providecommand \@ifnum [1]{%
 \ifnum #1\expandafter \@firstoftwo
 \else \expandafter \@secondoftwo
 \fi
}%
\providecommand \@ifx [1]{%
 \ifx #1\expandafter \@firstoftwo
 \else \expandafter \@secondoftwo
 \fi
}%
\providecommand \natexlab [1]{#1}%
\providecommand \enquote  [1]{``#1''}%
\providecommand \bibnamefont  [1]{#1}%
\providecommand \bibfnamefont [1]{#1}%
\providecommand \citenamefont [1]{#1}%
\providecommand \href@noop [0]{\@secondoftwo}%
\providecommand \href [0]{\begingroup \@sanitize@url \@href}%
\providecommand \@href[1]{\@@startlink{#1}\@@href}%
\providecommand \@@href[1]{\endgroup#1\@@endlink}%
\providecommand \@sanitize@url [0]{\catcode `\\12\catcode `\$12\catcode
  `\&12\catcode `\#12\catcode `\^12\catcode `\_12\catcode `\%12\relax}%
\providecommand \@@startlink[1]{}%
\providecommand \@@endlink[0]{}%
\providecommand \url  [0]{\begingroup\@sanitize@url \@url }%
\providecommand \@url [1]{\endgroup\@href {#1}{\urlprefix }}%
\providecommand \urlprefix  [0]{URL }%
\providecommand \Eprint [0]{\href }%
\providecommand \doibase [0]{https://doi.org/}%
\providecommand \selectlanguage [0]{\@gobble}%
\providecommand \bibinfo  [0]{\@secondoftwo}%
\providecommand \bibfield  [0]{\@secondoftwo}%
\providecommand \translation [1]{[#1]}%
\providecommand \BibitemOpen [0]{}%
\providecommand \bibitemStop [0]{}%
\providecommand \bibitemNoStop [0]{.\EOS\space}%
\providecommand \EOS [0]{\spacefactor3000\relax}%
\providecommand \BibitemShut  [1]{\csname bibitem#1\endcsname}%
\let\auto@bib@innerbib\@empty
%</preamble>
\bibitem [{\citenamefont {Hasan}\ and\ \citenamefont {Kane}(2010)}]{Hasan2010}%
  \BibitemOpen
  \bibfield  {author} {\bibinfo {author} {\bibfnamefont {M.~Z.}\ \bibnamefont
  {Hasan}}\ and\ \bibinfo {author} {\bibfnamefont {C.~L.}\ \bibnamefont
  {Kane}},\ }\bibfield  {title} {\bibinfo {title} {Colloquium: {Topological}
  insulators},\ }\href {https://doi.org/10.1103/RevModPhys.82.3045} {\bibfield
  {journal} {\bibinfo  {journal} {Rev. Mod. Phys.}\ }\textbf {\bibinfo {volume}
  {82}},\ \bibinfo {pages} {3045} (\bibinfo {year} {2010})}\BibitemShut
  {NoStop}%
\bibitem [{\citenamefont {von Klitzing}(2019)}]{von-Klitzing2019}%
  \BibitemOpen
  \bibfield  {author} {\bibinfo {author} {\bibfnamefont {K.}~\bibnamefont {von
  Klitzing}},\ }\bibfield  {title} {\bibinfo {title} {Essay: Quantum hall
  effect and the new international system of units},\ }\href
  {https://doi.org/10.1103/PhysRevLett.122.200001} {\bibfield  {journal}
  {\bibinfo  {journal} {Phys. Rev. Lett.}\ }\textbf {\bibinfo {volume} {122}},\
  \bibinfo {pages} {200001} (\bibinfo {year} {2019})}\BibitemShut {NoStop}%
\bibitem [{\citenamefont {Li}\ \emph {et~al.}(2014)\citenamefont {Li},
  \citenamefont {Xu},\ and\ \citenamefont {Chen}}]{Li2014}%
  \BibitemOpen
  \bibfield  {author} {\bibinfo {author} {\bibfnamefont {L.}~\bibnamefont
  {Li}}, \bibinfo {author} {\bibfnamefont {Z.}~\bibnamefont {Xu}},\ and\
  \bibinfo {author} {\bibfnamefont {S.}~\bibnamefont {Chen}},\ }\bibfield
  {title} {\bibinfo {title} {Topological phases of generalized
  {Su}-{Schrieffer}-{Heeger} models},\ }\href
  {https://doi.org/10.1103/PhysRevB.89.085111} {\bibfield  {journal} {\bibinfo
  {journal} {Phys. Rev. B}\ }\textbf {\bibinfo {volume} {89}},\ \bibinfo
  {pages} {085111} (\bibinfo {year} {2014})}\BibitemShut {NoStop}%
\bibitem [{\citenamefont {Zhao}\ \emph {et~al.}(2015)\citenamefont {Zhao},
  \citenamefont {Guo}, \citenamefont {Xia},\ and\ \citenamefont
  {Wang}}]{ZhaoGuoXiaWang2015-NanoPhotApp}%
  \BibitemOpen
  \bibfield  {author} {\bibinfo {author} {\bibfnamefont {H.}~\bibnamefont
  {Zhao}}, \bibinfo {author} {\bibfnamefont {Q.}~\bibnamefont {Guo}}, \bibinfo
  {author} {\bibfnamefont {F.}~\bibnamefont {Xia}},\ and\ \bibinfo {author}
  {\bibfnamefont {H.}~\bibnamefont {Wang}},\ }\bibfield  {title} {\bibinfo
  {title} {Two-dimensional materials for nanophotonics application},\ }\href
  {https://doi.org/doi:10.1515/nanoph-2014-0022} {\bibfield  {journal}
  {\bibinfo  {journal} {Nanophotonics}\ }\textbf {\bibinfo {volume} {4}},\
  \bibinfo {pages} {128} (\bibinfo {year} {2015})}\BibitemShut {NoStop}%
\bibitem [{\citenamefont {Gong}\ \emph {et~al.}(2018)\citenamefont {Gong},
  \citenamefont {Ashida}, \citenamefont {Kawabata}, \citenamefont {Takasan},
  \citenamefont {Higashikawa},\ and\ \citenamefont {Ueda}}]{Gong2018}%
  \BibitemOpen
  \bibfield  {author} {\bibinfo {author} {\bibfnamefont {Z.}~\bibnamefont
  {Gong}}, \bibinfo {author} {\bibfnamefont {Y.}~\bibnamefont {Ashida}},
  \bibinfo {author} {\bibfnamefont {K.}~\bibnamefont {Kawabata}}, \bibinfo
  {author} {\bibfnamefont {K.}~\bibnamefont {Takasan}}, \bibinfo {author}
  {\bibfnamefont {S.}~\bibnamefont {Higashikawa}},\ and\ \bibinfo {author}
  {\bibfnamefont {M.}~\bibnamefont {Ueda}},\ }\bibfield  {title} {\bibinfo
  {title} {Topological {Phases} of {Non}-{Hermitian} {Systems}},\ }\href
  {https://doi.org/10.1103/PhysRevX.8.031079} {\bibfield  {journal} {\bibinfo
  {journal} {Phys. Rev. X}\ }\textbf {\bibinfo {volume} {8}},\ \bibinfo {pages}
  {031079} (\bibinfo {year} {2018})}\BibitemShut {NoStop}%
\bibitem [{\citenamefont {Lindner}\ \emph {et~al.}(2011)\citenamefont
  {Lindner}, \citenamefont {Refael},\ and\ \citenamefont
  {Galitski}}]{Lindner2011}%
  \BibitemOpen
  \bibfield  {author} {\bibinfo {author} {\bibfnamefont {N.~H.}\ \bibnamefont
  {Lindner}}, \bibinfo {author} {\bibfnamefont {G.}~\bibnamefont {Refael}},\
  and\ \bibinfo {author} {\bibfnamefont {V.}~\bibnamefont {Galitski}},\
  }\bibfield  {title} {\bibinfo {title} {Floquet topological insulator in
  semiconductor quantum wells},\ }\href {https://doi.org/10.1038/nphys1926}
  {\bibfield  {journal} {\bibinfo  {journal} {Nat. Phys.}\ }\textbf {\bibinfo
  {volume} {7}},\ \bibinfo {pages} {490} (\bibinfo {year} {2011})}\BibitemShut
  {NoStop}%
\bibitem [{\citenamefont {G\'omez-Le\'on}\ and\ \citenamefont
  {Platero}(2013)}]{Gomez-Leon2013}%
  \BibitemOpen
  \bibfield  {author} {\bibinfo {author} {\bibfnamefont {A.}~\bibnamefont
  {G\'omez-Le\'on}}\ and\ \bibinfo {author} {\bibfnamefont {G.}~\bibnamefont
  {Platero}},\ }\bibfield  {title} {\bibinfo {title} {Floquet-{Bloch} {Theory}
  and {Topology} in {Periodically} {Driven} {Lattices}},\ }\href
  {https://doi.org/10.1103/PhysRevLett.110.200403} {\bibfield  {journal}
  {\bibinfo  {journal} {Phys. Rev. Lett.}\ }\textbf {\bibinfo {volume} {110}},\
  \bibinfo {pages} {200403} (\bibinfo {year} {2013})}\BibitemShut {NoStop}%
\bibitem [{\citenamefont {Harper}\ \emph {et~al.}(2020)\citenamefont {Harper},
  \citenamefont {Roy}, \citenamefont {Rudner},\ and\ \citenamefont
  {Sondhi}}]{harper_topology_2020}%
  \BibitemOpen
  \bibfield  {author} {\bibinfo {author} {\bibfnamefont {F.}~\bibnamefont
  {Harper}}, \bibinfo {author} {\bibfnamefont {R.}~\bibnamefont {Roy}},
  \bibinfo {author} {\bibfnamefont {M.~S.}\ \bibnamefont {Rudner}},\ and\
  \bibinfo {author} {\bibfnamefont {S.}~\bibnamefont {Sondhi}},\ }\bibfield
  {title} {\bibinfo {title} {Topology and {Broken} {Symmetry} in {Floquet}
  {Systems}},\ }\href
  {https://doi.org/10.1146/annurev-conmatphys-031218-013721} {\bibfield
  {journal} {\bibinfo  {journal} {Annu. Rev. Condens. Matter Phys.}\ }\textbf
  {\bibinfo {volume} {11}},\ \bibinfo {pages} {345} (\bibinfo {year}
  {2020})}\BibitemShut {NoStop}%
\bibitem [{\citenamefont {Rechtsman}\ \emph {et~al.}(2013)\citenamefont
  {Rechtsman}, \citenamefont {Zeuner}, \citenamefont {Plotnik}, \citenamefont
  {Lumer}, \citenamefont {Podolsky}, \citenamefont {Dreisow}, \citenamefont
  {Nolte}, \citenamefont {Segev},\ and\ \citenamefont
  {Szameit}}]{Rechtsman2013b}%
  \BibitemOpen
  \bibfield  {author} {\bibinfo {author} {\bibfnamefont {M.~C.}\ \bibnamefont
  {Rechtsman}}, \bibinfo {author} {\bibfnamefont {J.~M.}\ \bibnamefont
  {Zeuner}}, \bibinfo {author} {\bibfnamefont {Y.}~\bibnamefont {Plotnik}},
  \bibinfo {author} {\bibfnamefont {Y.}~\bibnamefont {Lumer}}, \bibinfo
  {author} {\bibfnamefont {D.}~\bibnamefont {Podolsky}}, \bibinfo {author}
  {\bibfnamefont {F.}~\bibnamefont {Dreisow}}, \bibinfo {author} {\bibfnamefont
  {S.}~\bibnamefont {Nolte}}, \bibinfo {author} {\bibfnamefont
  {M.}~\bibnamefont {Segev}},\ and\ \bibinfo {author} {\bibfnamefont
  {A.}~\bibnamefont {Szameit}},\ }\bibfield  {title} {\bibinfo {title}
  {Photonic {Floquet} topological insulators},\ }\href
  {https://doi.org/10.1038/nature12066} {\bibfield  {journal} {\bibinfo
  {journal} {Nature}\ }\textbf {\bibinfo {volume} {496}},\ \bibinfo {pages}
  {196} (\bibinfo {year} {2013})}\BibitemShut {NoStop}%
\bibitem [{\citenamefont {Fläschner}\ \emph {et~al.}(2016)\citenamefont
  {Fläschner}, \citenamefont {Rem}, \citenamefont {Tarnowski}, \citenamefont
  {Vogel}, \citenamefont {Lühmann}, \citenamefont {Sengstock},\ and\
  \citenamefont {Weitenberg}}]{Flaschner2016}%
  \BibitemOpen
  \bibfield  {author} {\bibinfo {author} {\bibfnamefont {N.}~\bibnamefont
  {Fläschner}}, \bibinfo {author} {\bibfnamefont {B.~S.}\ \bibnamefont {Rem}},
  \bibinfo {author} {\bibfnamefont {M.}~\bibnamefont {Tarnowski}}, \bibinfo
  {author} {\bibfnamefont {D.}~\bibnamefont {Vogel}}, \bibinfo {author}
  {\bibfnamefont {D.-S.}\ \bibnamefont {Lühmann}}, \bibinfo {author}
  {\bibfnamefont {K.}~\bibnamefont {Sengstock}},\ and\ \bibinfo {author}
  {\bibfnamefont {C.}~\bibnamefont {Weitenberg}},\ }\bibfield  {title}
  {\bibinfo {title} {Experimental reconstruction of the {Berry} curvature in a
  {Floquet} {Bloch} band},\ }\href {https://doi.org/10.1126/science.aad4568}
  {\bibfield  {journal} {\bibinfo  {journal} {Science}\ }\textbf {\bibinfo
  {volume} {352}},\ \bibinfo {pages} {1091} (\bibinfo {year}
  {2016})}\BibitemShut {NoStop}%
\bibitem [{\citenamefont {Kitagawa}\ \emph {et~al.}(2010)\citenamefont
  {Kitagawa}, \citenamefont {Berg}, \citenamefont {Rudner},\ and\ \citenamefont
  {Demler}}]{Kitagawa2010}%
  \BibitemOpen
  \bibfield  {author} {\bibinfo {author} {\bibfnamefont {T.}~\bibnamefont
  {Kitagawa}}, \bibinfo {author} {\bibfnamefont {E.}~\bibnamefont {Berg}},
  \bibinfo {author} {\bibfnamefont {M.}~\bibnamefont {Rudner}},\ and\ \bibinfo
  {author} {\bibfnamefont {E.}~\bibnamefont {Demler}},\ }\bibfield  {title}
  {\bibinfo {title} {Topological characterization of periodically driven
  quantum systems},\ }\href {https://doi.org/10.1103/PhysRevB.82.235114}
  {\bibfield  {journal} {\bibinfo  {journal} {Phys. Rev. B}\ }\textbf {\bibinfo
  {volume} {82}},\ \bibinfo {pages} {235114} (\bibinfo {year}
  {2010})}\BibitemShut {NoStop}%
\bibitem [{\citenamefont {Kitagawa}\ \emph {et~al.}(2012)\citenamefont
  {Kitagawa}, \citenamefont {Broome}, \citenamefont {Fedrizzi}, \citenamefont
  {Rudner}, \citenamefont {Berg}, \citenamefont {Kassal}, \citenamefont
  {Aspuru-Guzik}, \citenamefont {Demler},\ and\ \citenamefont
  {White}}]{Kitagawa2012}%
  \BibitemOpen
  \bibfield  {author} {\bibinfo {author} {\bibfnamefont {T.}~\bibnamefont
  {Kitagawa}}, \bibinfo {author} {\bibfnamefont {M.~A.}\ \bibnamefont
  {Broome}}, \bibinfo {author} {\bibfnamefont {A.}~\bibnamefont {Fedrizzi}},
  \bibinfo {author} {\bibfnamefont {M.~S.}\ \bibnamefont {Rudner}}, \bibinfo
  {author} {\bibfnamefont {E.}~\bibnamefont {Berg}}, \bibinfo {author}
  {\bibfnamefont {I.}~\bibnamefont {Kassal}}, \bibinfo {author} {\bibfnamefont
  {A.}~\bibnamefont {Aspuru-Guzik}}, \bibinfo {author} {\bibfnamefont
  {E.}~\bibnamefont {Demler}},\ and\ \bibinfo {author} {\bibfnamefont {A.~G.}\
  \bibnamefont {White}},\ }\bibfield  {title} {\bibinfo {title} {Observation of
  topologically protected bound states in photonic quantum walks},\ }\href
  {https://doi.org/10.1038/ncomms1872} {\bibfield  {journal} {\bibinfo
  {journal} {Nat. Commun.}\ }\textbf {\bibinfo {volume} {3}},\ \bibinfo {pages}
  {882} (\bibinfo {year} {2012})}\BibitemShut {NoStop}%
\bibitem [{\citenamefont {Rudner}\ \emph {et~al.}(2013)\citenamefont {Rudner},
  \citenamefont {Lindner}, \citenamefont {Berg},\ and\ \citenamefont
  {Levin}}]{Rudner2013}%
  \BibitemOpen
  \bibfield  {author} {\bibinfo {author} {\bibfnamefont {M.~S.}\ \bibnamefont
  {Rudner}}, \bibinfo {author} {\bibfnamefont {N.~H.}\ \bibnamefont {Lindner}},
  \bibinfo {author} {\bibfnamefont {E.}~\bibnamefont {Berg}},\ and\ \bibinfo
  {author} {\bibfnamefont {M.}~\bibnamefont {Levin}},\ }\bibfield  {title}
  {\bibinfo {title} {Anomalous {Edge} {States} and the {Bulk}-{Edge}
  {Correspondence} for {Periodically} {Driven} {Two}-{Dimensional} {Systems}},\
  }\href {https://doi.org/10.1103/PhysRevX.3.031005} {\bibfield  {journal}
  {\bibinfo  {journal} {Phys. Rev. X}\ }\textbf {\bibinfo {volume} {3}},\
  \bibinfo {pages} {031005} (\bibinfo {year} {2013})}\BibitemShut {NoStop}%
\bibitem [{\citenamefont {G\'omez-Le\'on}\ \emph {et~al.}(2014)\citenamefont
  {G\'omez-Le\'on}, \citenamefont {Delplace},\ and\ \citenamefont
  {Platero}}]{AGL2014}%
  \BibitemOpen
  \bibfield  {author} {\bibinfo {author} {\bibfnamefont {A.}~\bibnamefont
  {G\'omez-Le\'on}}, \bibinfo {author} {\bibfnamefont {P.}~\bibnamefont
  {Delplace}},\ and\ \bibinfo {author} {\bibfnamefont {G.}~\bibnamefont
  {Platero}},\ }\bibfield  {title} {\bibinfo {title} {Engineering anomalous
  quantum hall plateaus and antichiral states with ac fields},\ }\href
  {https://doi.org/10.1103/PhysRevB.89.205408} {\bibfield  {journal} {\bibinfo
  {journal} {Phys. Rev. B}\ }\textbf {\bibinfo {volume} {89}},\ \bibinfo
  {pages} {205408} (\bibinfo {year} {2014})}\BibitemShut {NoStop}%
\bibitem [{\citenamefont {Nathan}\ and\ \citenamefont
  {Rudner}(2015)}]{Nathan2015}%
  \BibitemOpen
  \bibfield  {author} {\bibinfo {author} {\bibfnamefont {F.}~\bibnamefont
  {Nathan}}\ and\ \bibinfo {author} {\bibfnamefont {M.~S.}\ \bibnamefont
  {Rudner}},\ }\bibfield  {title} {\bibinfo {title} {Topological singularities
  and the general classification of {Floquet}–{Bloch} systems},\ }\href
  {https://doi.org/10.1088/1367-2630/17/12/125014} {\bibfield  {journal}
  {\bibinfo  {journal} {New J. Phys.}\ }\textbf {\bibinfo {volume} {17}},\
  \bibinfo {pages} {125014} (\bibinfo {year} {2015})}\BibitemShut {NoStop}%
\bibitem [{\citenamefont {Peng}\ \emph {et~al.}(2016)\citenamefont {Peng},
  \citenamefont {Qin}, \citenamefont {Zhao}, \citenamefont {Shen},
  \citenamefont {Xu}, \citenamefont {Bao}, \citenamefont {Jia},\ and\
  \citenamefont {Zhu}}]{peng_experimental_2016}%
  \BibitemOpen
  \bibfield  {author} {\bibinfo {author} {\bibfnamefont {Y.-G.}\ \bibnamefont
  {Peng}}, \bibinfo {author} {\bibfnamefont {C.-Z.}\ \bibnamefont {Qin}},
  \bibinfo {author} {\bibfnamefont {D.-G.}\ \bibnamefont {Zhao}}, \bibinfo
  {author} {\bibfnamefont {Y.-X.}\ \bibnamefont {Shen}}, \bibinfo {author}
  {\bibfnamefont {X.-Y.}\ \bibnamefont {Xu}}, \bibinfo {author} {\bibfnamefont
  {M.}~\bibnamefont {Bao}}, \bibinfo {author} {\bibfnamefont {H.}~\bibnamefont
  {Jia}},\ and\ \bibinfo {author} {\bibfnamefont {X.-F.}\ \bibnamefont {Zhu}},\
  }\bibfield  {title} {\bibinfo {title} {Experimental demonstration of
  anomalous {Floquet} topological insulator for sound},\ }\href
  {https://doi.org/10.1038/ncomms13368} {\bibfield  {journal} {\bibinfo
  {journal} {Nat Commun}\ }\textbf {\bibinfo {volume} {7}},\ \bibinfo {pages}
  {13368} (\bibinfo {year} {2016})},\ \bibinfo {note} {peng2016}\BibitemShut
  {NoStop}%
\bibitem [{\citenamefont {Cheng}\ \emph {et~al.}(2019)\citenamefont {Cheng},
  \citenamefont {Pan}, \citenamefont {Wang}, \citenamefont {Zhang},
  \citenamefont {Yu}, \citenamefont {Gover}, \citenamefont {Zhang},
  \citenamefont {Li}, \citenamefont {Zhou},\ and\ \citenamefont
  {Zhu}}]{cheng_observation_2019}%
  \BibitemOpen
  \bibfield  {author} {\bibinfo {author} {\bibfnamefont {Q.}~\bibnamefont
  {Cheng}}, \bibinfo {author} {\bibfnamefont {Y.}~\bibnamefont {Pan}}, \bibinfo
  {author} {\bibfnamefont {H.}~\bibnamefont {Wang}}, \bibinfo {author}
  {\bibfnamefont {C.}~\bibnamefont {Zhang}}, \bibinfo {author} {\bibfnamefont
  {D.}~\bibnamefont {Yu}}, \bibinfo {author} {\bibfnamefont {A.}~\bibnamefont
  {Gover}}, \bibinfo {author} {\bibfnamefont {H.}~\bibnamefont {Zhang}},
  \bibinfo {author} {\bibfnamefont {T.}~\bibnamefont {Li}}, \bibinfo {author}
  {\bibfnamefont {L.}~\bibnamefont {Zhou}},\ and\ \bibinfo {author}
  {\bibfnamefont {S.}~\bibnamefont {Zhu}},\ }\bibfield  {title} {\bibinfo
  {title} {Observation of {Anomalous} $\pi$ {Modes} in {Photonic} {Floquet}
  {Engineering}},\ }\href {https://doi.org/10.1103/PhysRevLett.122.173901}
  {\bibfield  {journal} {\bibinfo  {journal} {Phys. Rev. Lett.}\ }\textbf
  {\bibinfo {volume} {122}},\ \bibinfo {pages} {173901} (\bibinfo {year}
  {2019})}\BibitemShut {NoStop}%
\bibitem [{\citenamefont {Afzal}\ \emph {et~al.}(2020)\citenamefont {Afzal},
  \citenamefont {Zimmerling}, \citenamefont {Ren}, \citenamefont {Perron},\
  and\ \citenamefont {Van}}]{afzal_realization_2020}%
  \BibitemOpen
  \bibfield  {author} {\bibinfo {author} {\bibfnamefont {S.}~\bibnamefont
  {Afzal}}, \bibinfo {author} {\bibfnamefont {T.~J.}\ \bibnamefont
  {Zimmerling}}, \bibinfo {author} {\bibfnamefont {Y.}~\bibnamefont {Ren}},
  \bibinfo {author} {\bibfnamefont {D.}~\bibnamefont {Perron}},\ and\ \bibinfo
  {author} {\bibfnamefont {V.}~\bibnamefont {Van}},\ }\bibfield  {title}
  {\bibinfo {title} {Realization of {Anomalous} {Floquet} {Insulators} in
  {Strongly} {Coupled} {Nanophotonic} {Lattices}},\ }\href
  {https://doi.org/10.1103/PhysRevLett.124.253601} {\bibfield  {journal}
  {\bibinfo  {journal} {Phys. Rev. Lett.}\ }\textbf {\bibinfo {volume} {124}},\
  \bibinfo {pages} {253601} (\bibinfo {year} {2020})}\BibitemShut {NoStop}%
\bibitem [{\citenamefont {Wintersperger}\ \emph {et~al.}(2020)\citenamefont
  {Wintersperger}, \citenamefont {Braun}, \citenamefont {Ünal}, \citenamefont
  {Eckardt}, \citenamefont {Liberto}, \citenamefont {Goldman}, \citenamefont
  {Bloch},\ and\ \citenamefont {Aidelsburger}}]{Wintersperger2020}%
  \BibitemOpen
  \bibfield  {author} {\bibinfo {author} {\bibfnamefont {K.}~\bibnamefont
  {Wintersperger}}, \bibinfo {author} {\bibfnamefont {C.}~\bibnamefont
  {Braun}}, \bibinfo {author} {\bibfnamefont {F.~N.}\ \bibnamefont {Ünal}},
  \bibinfo {author} {\bibfnamefont {A.}~\bibnamefont {Eckardt}}, \bibinfo
  {author} {\bibfnamefont {M.~D.}\ \bibnamefont {Liberto}}, \bibinfo {author}
  {\bibfnamefont {N.}~\bibnamefont {Goldman}}, \bibinfo {author} {\bibfnamefont
  {I.}~\bibnamefont {Bloch}},\ and\ \bibinfo {author} {\bibfnamefont
  {M.}~\bibnamefont {Aidelsburger}},\ }\bibfield  {title} {\bibinfo {title}
  {Realization of an anomalous {Floquet} topological system with ultracold
  atoms},\ }\href {https://doi.org/10.1038/s41567-020-0949-y} {\bibfield
  {journal} {\bibinfo  {journal} {Nat. Phys.}\ }\textbf {\bibinfo {volume}
  {16}},\ \bibinfo {pages} {1058} (\bibinfo {year} {2020})}\BibitemShut
  {NoStop}%
\bibitem [{\citenamefont {G\'omez-Le\'on}(2023)}]{AGL-Floquet-2023}%
  \BibitemOpen
  \bibfield  {author} {\bibinfo {author} {\bibfnamefont {A.}~\bibnamefont
  {G\'omez-Le\'on}},\ }\href@noop {} {\bibinfo {title} {Anomalous floquet
  phases. a resonance phenomena}} (\bibinfo {year} {2023})\BibitemShut
  {NoStop}%
\bibitem [{\citenamefont {Zhang}\ \emph {et~al.}(2023)\citenamefont {Zhang},
  \citenamefont {Yi}, \citenamefont {Zhang}, \citenamefont {Jiao},
  \citenamefont {Shi}, \citenamefont {Yuan}, \citenamefont {Zhang},
  \citenamefont {Liu}, \citenamefont {Chen},\ and\ \citenamefont
  {Pan}}]{zhang_tuning_2023}%
  \BibitemOpen
  \bibfield  {author} {\bibinfo {author} {\bibfnamefont {J.-Y.}\ \bibnamefont
  {Zhang}}, \bibinfo {author} {\bibfnamefont {C.-R.}\ \bibnamefont {Yi}},
  \bibinfo {author} {\bibfnamefont {L.}~\bibnamefont {Zhang}}, \bibinfo
  {author} {\bibfnamefont {R.-H.}\ \bibnamefont {Jiao}}, \bibinfo {author}
  {\bibfnamefont {K.-Y.}\ \bibnamefont {Shi}}, \bibinfo {author} {\bibfnamefont
  {H.}~\bibnamefont {Yuan}}, \bibinfo {author} {\bibfnamefont {W.}~\bibnamefont
  {Zhang}}, \bibinfo {author} {\bibfnamefont {X.-J.}\ \bibnamefont {Liu}},
  \bibinfo {author} {\bibfnamefont {S.}~\bibnamefont {Chen}},\ and\ \bibinfo
  {author} {\bibfnamefont {J.-W.}\ \bibnamefont {Pan}},\ }\bibfield  {title}
  {\bibinfo {title} {Tuning {Anomalous} {Floquet} {Topological} {Bands} with
  {Ultracold} {Atoms}},\ }\href
  {https://doi.org/10.1103/PhysRevLett.130.043201} {\bibfield  {journal}
  {\bibinfo  {journal} {Phys. Rev. Lett.}\ }\textbf {\bibinfo {volume} {130}},\
  \bibinfo {pages} {043201} (\bibinfo {year} {2023})}\BibitemShut {NoStop}%
\bibitem [{\citenamefont {Braun}\ \emph {et~al.}(2023)\citenamefont {Braun},
  \citenamefont {Saint-Jalm}, \citenamefont {Hesse}, \citenamefont {Arceri},
  \citenamefont {Bloch},\ and\ \citenamefont
  {Aidelsburger}}]{braun_real-space_2023}%
  \BibitemOpen
  \bibfield  {author} {\bibinfo {author} {\bibfnamefont {C.}~\bibnamefont
  {Braun}}, \bibinfo {author} {\bibfnamefont {R.}~\bibnamefont {Saint-Jalm}},
  \bibinfo {author} {\bibfnamefont {A.}~\bibnamefont {Hesse}}, \bibinfo
  {author} {\bibfnamefont {J.}~\bibnamefont {Arceri}}, \bibinfo {author}
  {\bibfnamefont {I.}~\bibnamefont {Bloch}},\ and\ \bibinfo {author}
  {\bibfnamefont {M.}~\bibnamefont {Aidelsburger}},\ }\href
  {https://doi.org/10.48550/arXiv.2304.01980} {\bibinfo {title} {Real-space
  detection and manipulation of topological edge modes with ultracold atoms}}
  (\bibinfo {year} {2023}),\ \bibinfo {note} {arXiv:2304.01980}\BibitemShut
  {NoStop}%
\bibitem [{\citenamefont {Asb\'oth}\ and\ \citenamefont
  {Obuse}(2013)}]{Asboth2013}%
  \BibitemOpen
  \bibfield  {author} {\bibinfo {author} {\bibfnamefont {J.~K.}\ \bibnamefont
  {Asb\'oth}}\ and\ \bibinfo {author} {\bibfnamefont {H.}~\bibnamefont
  {Obuse}},\ }\bibfield  {title} {\bibinfo {title} {Bulk-boundary
  correspondence for chiral symmetric quantum walks},\ }\href
  {https://doi.org/10.1103/PhysRevB.88.121406} {\bibfield  {journal} {\bibinfo
  {journal} {Phys. Rev. B}\ }\textbf {\bibinfo {volume} {88}},\ \bibinfo
  {pages} {121406(R)} (\bibinfo {year} {2013})}\BibitemShut {NoStop}%
\bibitem [{\citenamefont {Barkhofen}\ \emph {et~al.}(2017)\citenamefont
  {Barkhofen}, \citenamefont {Nitsche}, \citenamefont {Elster}, \citenamefont
  {Lorz}, \citenamefont {Gábris}, \citenamefont {Jex},\ and\ \citenamefont
  {Silberhorn}}]{barkhofen_measuring_2017}%
  \BibitemOpen
  \bibfield  {author} {\bibinfo {author} {\bibfnamefont {S.}~\bibnamefont
  {Barkhofen}}, \bibinfo {author} {\bibfnamefont {T.}~\bibnamefont {Nitsche}},
  \bibinfo {author} {\bibfnamefont {F.}~\bibnamefont {Elster}}, \bibinfo
  {author} {\bibfnamefont {L.}~\bibnamefont {Lorz}}, \bibinfo {author}
  {\bibfnamefont {A.}~\bibnamefont {Gábris}}, \bibinfo {author} {\bibfnamefont
  {I.}~\bibnamefont {Jex}},\ and\ \bibinfo {author} {\bibfnamefont
  {C.}~\bibnamefont {Silberhorn}},\ }\bibfield  {title} {\bibinfo {title}
  {Measuring topological invariants in disordered discrete-time quantum
  walks},\ }\href {https://doi.org/10.1103/PhysRevA.96.033846} {\bibfield
  {journal} {\bibinfo  {journal} {Phys. Rev. A}\ }\textbf {\bibinfo {volume}
  {96}},\ \bibinfo {pages} {033846} (\bibinfo {year} {2017})}\BibitemShut
  {NoStop}%
\bibitem [{\citenamefont {Maczewsky}\ \emph {et~al.}(2017)\citenamefont
  {Maczewsky}, \citenamefont {Zeuner}, \citenamefont {Nolte},\ and\
  \citenamefont {Szameit}}]{Maczewsky2017}%
  \BibitemOpen
  \bibfield  {author} {\bibinfo {author} {\bibfnamefont {L.~J.}\ \bibnamefont
  {Maczewsky}}, \bibinfo {author} {\bibfnamefont {J.~M.}\ \bibnamefont
  {Zeuner}}, \bibinfo {author} {\bibfnamefont {S.}~\bibnamefont {Nolte}},\ and\
  \bibinfo {author} {\bibfnamefont {A.}~\bibnamefont {Szameit}},\ }\bibfield
  {title} {\bibinfo {title} {Observation of photonic anomalous {Floquet}
  topological insulators},\ }\href {https://doi.org/10.1038/ncomms13756}
  {\bibfield  {journal} {\bibinfo  {journal} {Nat. Commun.}\ }\textbf {\bibinfo
  {volume} {8}},\ \bibinfo {pages} {13756} (\bibinfo {year}
  {2017})}\BibitemShut {NoStop}%
\bibitem [{\citenamefont {Mukherjee}\ \emph {et~al.}(2017)\citenamefont
  {Mukherjee}, \citenamefont {Spracklen}, \citenamefont {Valiente},
  \citenamefont {Andersson}, \citenamefont {Öhberg}, \citenamefont {Goldman},\
  and\ \citenamefont {Thomson}}]{Mukherjee2017a}%
  \BibitemOpen
  \bibfield  {author} {\bibinfo {author} {\bibfnamefont {S.}~\bibnamefont
  {Mukherjee}}, \bibinfo {author} {\bibfnamefont {A.}~\bibnamefont
  {Spracklen}}, \bibinfo {author} {\bibfnamefont {M.}~\bibnamefont {Valiente}},
  \bibinfo {author} {\bibfnamefont {E.}~\bibnamefont {Andersson}}, \bibinfo
  {author} {\bibfnamefont {P.}~\bibnamefont {Öhberg}}, \bibinfo {author}
  {\bibfnamefont {N.}~\bibnamefont {Goldman}},\ and\ \bibinfo {author}
  {\bibfnamefont {R.~R.}\ \bibnamefont {Thomson}},\ }\bibfield  {title}
  {\bibinfo {title} {Experimental observation of anomalous topological edge
  modes in a slowly driven photonic lattice},\ }\href
  {https://doi.org/10.1038/ncomms13918} {\bibfield  {journal} {\bibinfo
  {journal} {Nat. Commun.}\ }\textbf {\bibinfo {volume} {8}},\ \bibinfo {pages}
  {13918} (\bibinfo {year} {2017})}\BibitemShut {NoStop}%
\bibitem [{\citenamefont {Bellec}\ \emph {et~al.}(2017)\citenamefont {Bellec},
  \citenamefont {Michel}, \citenamefont {Zhang}, \citenamefont {Tzortzakis},\
  and\ \citenamefont {Delplace}}]{Bellec2017}%
  \BibitemOpen
  \bibfield  {author} {\bibinfo {author} {\bibfnamefont {M.}~\bibnamefont
  {Bellec}}, \bibinfo {author} {\bibfnamefont {C.}~\bibnamefont {Michel}},
  \bibinfo {author} {\bibfnamefont {H.}~\bibnamefont {Zhang}}, \bibinfo
  {author} {\bibfnamefont {S.}~\bibnamefont {Tzortzakis}},\ and\ \bibinfo
  {author} {\bibfnamefont {P.}~\bibnamefont {Delplace}},\ }\bibfield  {title}
  {\bibinfo {title} {Non-diffracting states in one-dimensional {Floquet}
  photonic topological insulators},\ }\href
  {https://doi.org/10.1209/0295-5075/119/14003} {\bibfield  {journal} {\bibinfo
   {journal} {EPL}\ }\textbf {\bibinfo {volume} {119}},\ \bibinfo {pages}
  {14003} (\bibinfo {year} {2017})}\BibitemShut {NoStop}%
\bibitem [{\citenamefont {Bisianov}\ \emph {et~al.}(2019)\citenamefont
  {Bisianov}, \citenamefont {Wimmer}, \citenamefont {Peschel},\ and\
  \citenamefont {Egorov}}]{Bisianov2019}%
  \BibitemOpen
  \bibfield  {author} {\bibinfo {author} {\bibfnamefont {A.}~\bibnamefont
  {Bisianov}}, \bibinfo {author} {\bibfnamefont {M.}~\bibnamefont {Wimmer}},
  \bibinfo {author} {\bibfnamefont {U.}~\bibnamefont {Peschel}},\ and\ \bibinfo
  {author} {\bibfnamefont {O.~A.}\ \bibnamefont {Egorov}},\ }\bibfield  {title}
  {\bibinfo {title} {Stability of topologically protected edge states in
  nonlinear fiber loops},\ }\href {https://doi.org/10.1103/PhysRevA.100.063830}
  {\bibfield  {journal} {\bibinfo  {journal} {Phys. Rev. A}\ }\textbf {\bibinfo
  {volume} {100}},\ \bibinfo {pages} {063830} (\bibinfo {year}
  {2019})}\BibitemShut {NoStop}%
\bibitem [{\citenamefont {Cedzich}\ \emph {et~al.}(2018)\citenamefont
  {Cedzich}, \citenamefont {Geib}, \citenamefont {Grünbaum}, \citenamefont
  {Stahl}, \citenamefont {Velázquez}, \citenamefont {Werner},\ and\
  \citenamefont {Werner}}]{Cedzich2018}%
  \BibitemOpen
  \bibfield  {author} {\bibinfo {author} {\bibfnamefont {C.}~\bibnamefont
  {Cedzich}}, \bibinfo {author} {\bibfnamefont {T.}~\bibnamefont {Geib}},
  \bibinfo {author} {\bibfnamefont {F.~A.}\ \bibnamefont {Grünbaum}}, \bibinfo
  {author} {\bibfnamefont {C.}~\bibnamefont {Stahl}}, \bibinfo {author}
  {\bibfnamefont {L.}~\bibnamefont {Velázquez}}, \bibinfo {author}
  {\bibfnamefont {A.~H.}\ \bibnamefont {Werner}},\ and\ \bibinfo {author}
  {\bibfnamefont {R.~F.}\ \bibnamefont {Werner}},\ }\bibfield  {title}
  {\bibinfo {title} {The {Topological} {Classification} of {One}-{Dimensional}
  {Symmetric} {Quantum} {Walks}},\ }\href
  {https://doi.org/10.1007/s00023-017-0630-x} {\bibfield  {journal} {\bibinfo
  {journal} {Ann. Henri Poincaré}\ }\textbf {\bibinfo {volume} {19}},\
  \bibinfo {pages} {325} (\bibinfo {year} {2018})}\BibitemShut {NoStop}%
\bibitem [{\citenamefont {Mochizuki}\ \emph {et~al.}(2020)\citenamefont
  {Mochizuki}, \citenamefont {Bessho}, \citenamefont {Sato},\ and\
  \citenamefont {Obuse}}]{Mochizuki2020}%
  \BibitemOpen
  \bibfield  {author} {\bibinfo {author} {\bibfnamefont {K.}~\bibnamefont
  {Mochizuki}}, \bibinfo {author} {\bibfnamefont {T.}~\bibnamefont {Bessho}},
  \bibinfo {author} {\bibfnamefont {M.}~\bibnamefont {Sato}},\ and\ \bibinfo
  {author} {\bibfnamefont {H.}~\bibnamefont {Obuse}},\ }\bibfield  {title}
  {\bibinfo {title} {Topological quantum walk with discrete time-glide
  symmetry},\ }\href {https://doi.org/10.1103/PhysRevB.102.035418} {\bibfield
  {journal} {\bibinfo  {journal} {Phys. Rev. B}\ }\textbf {\bibinfo {volume}
  {102}},\ \bibinfo {pages} {035418} (\bibinfo {year} {2020})}\BibitemShut
  {NoStop}%
\bibitem [{\citenamefont {Grudka}\ \emph {et~al.}(2023)\citenamefont {Grudka},
  \citenamefont {Karczewski}, \citenamefont {Kurzy\ifmmode~\acute{n}\else
  \'{n}\fi{}ski}, \citenamefont {W\'ojcik},\ and\ \citenamefont
  {W\'ojcik}}]{Grudka2023}%
  \BibitemOpen
  \bibfield  {author} {\bibinfo {author} {\bibfnamefont {A.}~\bibnamefont
  {Grudka}}, \bibinfo {author} {\bibfnamefont {M.}~\bibnamefont {Karczewski}},
  \bibinfo {author} {\bibfnamefont {P.}~\bibnamefont
  {Kurzy\ifmmode~\acute{n}\else \'{n}\fi{}ski}}, \bibinfo {author}
  {\bibfnamefont {J.}~\bibnamefont {W\'ojcik}},\ and\ \bibinfo {author}
  {\bibfnamefont {A.}~\bibnamefont {W\'ojcik}},\ }\bibfield  {title} {\bibinfo
  {title} {Topological invariants in quantum walks},\ }\href
  {https://doi.org/10.1103/PhysRevA.107.032201} {\bibfield  {journal} {\bibinfo
   {journal} {Phys. Rev. A}\ }\textbf {\bibinfo {volume} {107}},\ \bibinfo
  {pages} {032201} (\bibinfo {year} {2023})}\BibitemShut {NoStop}%
\bibitem [{\citenamefont {Bessho}\ \emph {et~al.}(2022)\citenamefont {Bessho},
  \citenamefont {Mochizuki}, \citenamefont {Obuse},\ and\ \citenamefont
  {Sato}}]{Bessho2022}%
  \BibitemOpen
  \bibfield  {author} {\bibinfo {author} {\bibfnamefont {T.}~\bibnamefont
  {Bessho}}, \bibinfo {author} {\bibfnamefont {K.}~\bibnamefont {Mochizuki}},
  \bibinfo {author} {\bibfnamefont {H.}~\bibnamefont {Obuse}},\ and\ \bibinfo
  {author} {\bibfnamefont {M.}~\bibnamefont {Sato}},\ }\bibfield  {title}
  {\bibinfo {title} {Extrinsic topology of floquet anomalous boundary states in
  quantum walks},\ }\href {https://doi.org/10.1103/PhysRevB.105.094306}
  {\bibfield  {journal} {\bibinfo  {journal} {Phys. Rev. B}\ }\textbf {\bibinfo
  {volume} {105}},\ \bibinfo {pages} {094306} (\bibinfo {year}
  {2022})}\BibitemShut {NoStop}%
\bibitem [{\citenamefont {Mukherjee}\ and\ \citenamefont
  {Rechtsman}(2020)}]{Mukherjee2020}%
  \BibitemOpen
  \bibfield  {author} {\bibinfo {author} {\bibfnamefont {S.}~\bibnamefont
  {Mukherjee}}\ and\ \bibinfo {author} {\bibfnamefont {M.~C.}\ \bibnamefont
  {Rechtsman}},\ }\bibfield  {title} {\bibinfo {title} {Observation of
  {Floquet} solitons in a topological bandgap},\ }\href
  {https://doi.org/10.1126/science.aba8725} {\bibfield  {journal} {\bibinfo
  {journal} {Science}\ }\textbf {\bibinfo {volume} {368}},\ \bibinfo {pages}
  {856 LP } (\bibinfo {year} {2020})}\BibitemShut {NoStop}%
\bibitem [{\citenamefont {Mukherjee}\ and\ \citenamefont
  {Rechtsman}(2021)}]{Mukherjee2021}%
  \BibitemOpen
  \bibfield  {author} {\bibinfo {author} {\bibfnamefont {S.}~\bibnamefont
  {Mukherjee}}\ and\ \bibinfo {author} {\bibfnamefont {M.~C.}\ \bibnamefont
  {Rechtsman}},\ }\bibfield  {title} {\bibinfo {title} {Observation of
  {Unidirectional} {Solitonlike} {Edge} {States} in {Nonlinear} {Floquet}
  {Topological} {Insulators}},\ }\href
  {https://doi.org/10.1103/PhysRevX.11.041057} {\bibfield  {journal} {\bibinfo
  {journal} {Phys. Rev. X}\ }\textbf {\bibinfo {volume} {11}},\ \bibinfo
  {pages} {041057} (\bibinfo {year} {2021})}\BibitemShut {NoStop}%
\bibitem [{\citenamefont {Wimmer}\ \emph
  {et~al.}(2015{\natexlab{a}})\citenamefont {Wimmer}, \citenamefont {Miri},
  \citenamefont {Christodoulides},\ and\ \citenamefont {Peschel}}]{Wimmer2015}%
  \BibitemOpen
  \bibfield  {author} {\bibinfo {author} {\bibfnamefont {M.}~\bibnamefont
  {Wimmer}}, \bibinfo {author} {\bibfnamefont {M.-A.}\ \bibnamefont {Miri}},
  \bibinfo {author} {\bibfnamefont {D.}~\bibnamefont {Christodoulides}},\ and\
  \bibinfo {author} {\bibfnamefont {U.}~\bibnamefont {Peschel}},\ }\bibfield
  {title} {\bibinfo {title} {Observation of {Bloch} oscillations in complex
  {PT}-symmetric photonic lattices},\ }\href
  {http://dx.doi.org/10.1038/srep17760} {\bibfield  {journal} {\bibinfo
  {journal} {Sci. Rep.}\ }\textbf {\bibinfo {volume} {5}},\ \bibinfo {pages}
  {17760} (\bibinfo {year} {2015}{\natexlab{a}})}\BibitemShut {NoStop}%
\bibitem [{\citenamefont {Muniz}\ \emph {et~al.}(2019)\citenamefont {Muniz},
  \citenamefont {Wimmer}, \citenamefont {Bisianov}, \citenamefont {Peschel},
  \citenamefont {Morandotti}, \citenamefont {Jung},\ and\ \citenamefont
  {Christodoulides}}]{Muniz2019}%
  \BibitemOpen
  \bibfield  {author} {\bibinfo {author} {\bibfnamefont {A.}~\bibnamefont
  {Muniz}}, \bibinfo {author} {\bibfnamefont {M.}~\bibnamefont {Wimmer}},
  \bibinfo {author} {\bibfnamefont {A.}~\bibnamefont {Bisianov}}, \bibinfo
  {author} {\bibfnamefont {U.}~\bibnamefont {Peschel}}, \bibinfo {author}
  {\bibfnamefont {R.}~\bibnamefont {Morandotti}}, \bibinfo {author}
  {\bibfnamefont {P.~S.}\ \bibnamefont {Jung}},\ and\ \bibinfo {author}
  {\bibfnamefont {D.~N.}\ \bibnamefont {Christodoulides}},\ }\bibfield  {title}
  {\bibinfo {title} {{2D} {Solitons} in {P} {T} -{Symmetric} {Photonic}
  {Lattices}},\ }\href {https://doi.org/10.1103/PhysRevLett.123.253903}
  {\bibfield  {journal} {\bibinfo  {journal} {Phys. Rev. Lett.}\ }\textbf
  {\bibinfo {volume} {123}},\ \bibinfo {pages} {253903} (\bibinfo {year}
  {2019})}\BibitemShut {NoStop}%
\bibitem [{\citenamefont {Wimmer}\ \emph
  {et~al.}(2015{\natexlab{b}})\citenamefont {Wimmer}, \citenamefont
  {Regensburger}, \citenamefont {Miri}, \citenamefont {Bersch}, \citenamefont
  {Christodoulides},\ and\ \citenamefont {Peschel}}]{Wimmer2015a}%
  \BibitemOpen
  \bibfield  {author} {\bibinfo {author} {\bibfnamefont {M.}~\bibnamefont
  {Wimmer}}, \bibinfo {author} {\bibfnamefont {A.}~\bibnamefont
  {Regensburger}}, \bibinfo {author} {\bibfnamefont {M.-A.}\ \bibnamefont
  {Miri}}, \bibinfo {author} {\bibfnamefont {C.}~\bibnamefont {Bersch}},
  \bibinfo {author} {\bibfnamefont {D.~N.}\ \bibnamefont {Christodoulides}},\
  and\ \bibinfo {author} {\bibfnamefont {U.}~\bibnamefont {Peschel}},\
  }\bibfield  {title} {\bibinfo {title} {Observation of optical solitons in
  {PT}-symmetric lattices},\ }\href {https://doi.org/10.1038/ncomms8782}
  {\bibfield  {journal} {\bibinfo  {journal} {Nat. Commun.}\ }\textbf {\bibinfo
  {volume} {6}},\ \bibinfo {pages} {7782} (\bibinfo {year}
  {2015}{\natexlab{b}})}\BibitemShut {NoStop}%
\bibitem [{\citenamefont {Chalabi}\ \emph {et~al.}(2019)\citenamefont
  {Chalabi}, \citenamefont {Barik}, \citenamefont {Mittal}, \citenamefont
  {Murphy}, \citenamefont {Hafezi},\ and\ \citenamefont {Waks}}]{Chalabi2019}%
  \BibitemOpen
  \bibfield  {author} {\bibinfo {author} {\bibfnamefont {H.}~\bibnamefont
  {Chalabi}}, \bibinfo {author} {\bibfnamefont {S.}~\bibnamefont {Barik}},
  \bibinfo {author} {\bibfnamefont {S.}~\bibnamefont {Mittal}}, \bibinfo
  {author} {\bibfnamefont {T.~E.}\ \bibnamefont {Murphy}}, \bibinfo {author}
  {\bibfnamefont {M.}~\bibnamefont {Hafezi}},\ and\ \bibinfo {author}
  {\bibfnamefont {E.}~\bibnamefont {Waks}},\ }\bibfield  {title} {\bibinfo
  {title} {Synthetic {Gauge} {Field} for {Two}-{Dimensional}
  {Time}-{Multiplexed} {Quantum} {Random} {Walks}},\ }\href
  {https://doi.org/10.1103/PhysRevLett.123.150503} {\bibfield  {journal}
  {\bibinfo  {journal} {Phys. Rev. Lett.}\ }\textbf {\bibinfo {volume} {123}},\
  \bibinfo {pages} {150503} (\bibinfo {year} {2019})}\BibitemShut {NoStop}%
\bibitem [{\citenamefont {Weidemann}\ \emph {et~al.}(2020)\citenamefont
  {Weidemann}, \citenamefont {Kremer}, \citenamefont {Helbig}, \citenamefont
  {Hofmann}, \citenamefont {Stegmaier}, \citenamefont {Greiter}, \citenamefont
  {Thomale},\ and\ \citenamefont {Szameit}}]{Weidemann2020}%
  \BibitemOpen
  \bibfield  {author} {\bibinfo {author} {\bibfnamefont {S.}~\bibnamefont
  {Weidemann}}, \bibinfo {author} {\bibfnamefont {M.}~\bibnamefont {Kremer}},
  \bibinfo {author} {\bibfnamefont {T.}~\bibnamefont {Helbig}}, \bibinfo
  {author} {\bibfnamefont {T.}~\bibnamefont {Hofmann}}, \bibinfo {author}
  {\bibfnamefont {A.}~\bibnamefont {Stegmaier}}, \bibinfo {author}
  {\bibfnamefont {M.}~\bibnamefont {Greiter}}, \bibinfo {author} {\bibfnamefont
  {R.}~\bibnamefont {Thomale}},\ and\ \bibinfo {author} {\bibfnamefont
  {A.}~\bibnamefont {Szameit}},\ }\bibfield  {title} {\bibinfo {title}
  {Topological funneling of light},\ }\href
  {https://doi.org/10.1126/science.aaz8727} {\bibfield  {journal} {\bibinfo
  {journal} {Science}\ }\textbf {\bibinfo {volume} {368}},\ \bibinfo {pages}
  {311} (\bibinfo {year} {2020})}\BibitemShut {NoStop}%
\bibitem [{\citenamefont {Wimmer}\ \emph {et~al.}(2021)\citenamefont {Wimmer},
  \citenamefont {Monika}, \citenamefont {Carusotto}, \citenamefont {Peschel},\
  and\ \citenamefont {Price}}]{Wimmer2021}%
  \BibitemOpen
  \bibfield  {author} {\bibinfo {author} {\bibfnamefont {M.}~\bibnamefont
  {Wimmer}}, \bibinfo {author} {\bibfnamefont {M.}~\bibnamefont {Monika}},
  \bibinfo {author} {\bibfnamefont {I.}~\bibnamefont {Carusotto}}, \bibinfo
  {author} {\bibfnamefont {U.}~\bibnamefont {Peschel}},\ and\ \bibinfo {author}
  {\bibfnamefont {H.~M.}\ \bibnamefont {Price}},\ }\bibfield  {title} {\bibinfo
  {title} {Superfluidity of {Light} and {Its} {Breakdown} in {Optical} {Mesh}
  {Lattices}},\ }\href {https://doi.org/10.1103/PhysRevLett.127.163901}
  {\bibfield  {journal} {\bibinfo  {journal} {Phys. Rev. Lett.}\ }\textbf
  {\bibinfo {volume} {127}},\ \bibinfo {pages} {163901} (\bibinfo {year}
  {2021})}\BibitemShut {NoStop}%
\bibitem [{\citenamefont {Adiyatullin}\ \emph {et~al.}(2023)\citenamefont
  {Adiyatullin}, \citenamefont {Upreti}, \citenamefont {Lechevalier},
  \citenamefont {Evain}, \citenamefont {Copie}, \citenamefont {Suret},
  \citenamefont {Randoux}, \citenamefont {Delplace},\ and\ \citenamefont
  {Amo}}]{Adiyatullin2023}%
  \BibitemOpen
  \bibfield  {author} {\bibinfo {author} {\bibfnamefont {A.~F.}\ \bibnamefont
  {Adiyatullin}}, \bibinfo {author} {\bibfnamefont {L.~K.}\ \bibnamefont
  {Upreti}}, \bibinfo {author} {\bibfnamefont {C.}~\bibnamefont {Lechevalier}},
  \bibinfo {author} {\bibfnamefont {C.}~\bibnamefont {Evain}}, \bibinfo
  {author} {\bibfnamefont {F.}~\bibnamefont {Copie}}, \bibinfo {author}
  {\bibfnamefont {P.}~\bibnamefont {Suret}}, \bibinfo {author} {\bibfnamefont
  {S.}~\bibnamefont {Randoux}}, \bibinfo {author} {\bibfnamefont
  {P.}~\bibnamefont {Delplace}},\ and\ \bibinfo {author} {\bibfnamefont
  {A.}~\bibnamefont {Amo}},\ }\bibfield  {title} {\bibinfo {title} {Topological
  {Properties} of {Floquet} {Winding} {Bands} in a {Photonic} {Lattice}},\
  }\href {https://doi.org/10.1103/PhysRevLett.130.056901} {\bibfield  {journal}
  {\bibinfo  {journal} {Phys. Rev. Lett.}\ }\textbf {\bibinfo {volume} {130}},\
  \bibinfo {pages} {056901} (\bibinfo {year} {2023})}\BibitemShut {NoStop}%
\bibitem [{\citenamefont {Lechevalier}\ \emph {et~al.}(2021)\citenamefont
  {Lechevalier}, \citenamefont {Evain}, \citenamefont {Suret}, \citenamefont
  {Copie}, \citenamefont {Amo},\ and\ \citenamefont
  {Randoux}}]{Lechevalier2021}%
  \BibitemOpen
  \bibfield  {author} {\bibinfo {author} {\bibfnamefont {C.}~\bibnamefont
  {Lechevalier}}, \bibinfo {author} {\bibfnamefont {C.}~\bibnamefont {Evain}},
  \bibinfo {author} {\bibfnamefont {P.}~\bibnamefont {Suret}}, \bibinfo
  {author} {\bibfnamefont {F.}~\bibnamefont {Copie}}, \bibinfo {author}
  {\bibfnamefont {A.}~\bibnamefont {Amo}},\ and\ \bibinfo {author}
  {\bibfnamefont {S.}~\bibnamefont {Randoux}},\ }\bibfield  {title} {\bibinfo
  {title} {Single-shot measurement of the photonic band structure in a
  fiber-based {Floquet}-{Bloch} lattice},\ }\href
  {https://doi.org/10.1038/s42005-021-00750-w} {\bibfield  {journal} {\bibinfo
  {journal} {Commun. Phys.}\ }\textbf {\bibinfo {volume} {4}},\ \bibinfo
  {pages} {243} (\bibinfo {year} {2021})}\BibitemShut {NoStop}%
\bibitem [{\citenamefont {Wimmer}\ \emph {et~al.}(2017)\citenamefont {Wimmer},
  \citenamefont {Price}, \citenamefont {Carusotto},\ and\ \citenamefont
  {Peschel}}]{Wimmer2017}%
  \BibitemOpen
  \bibfield  {author} {\bibinfo {author} {\bibfnamefont {M.}~\bibnamefont
  {Wimmer}}, \bibinfo {author} {\bibfnamefont {H.~M.}\ \bibnamefont {Price}},
  \bibinfo {author} {\bibfnamefont {I.}~\bibnamefont {Carusotto}},\ and\
  \bibinfo {author} {\bibfnamefont {U.}~\bibnamefont {Peschel}},\ }\bibfield
  {title} {\bibinfo {title} {Experimental measurement of the {Berry} curvature
  from anomalous transport},\ }\href {http://dx.doi.org/10.1038/nphys4050}
  {\bibfield  {journal} {\bibinfo  {journal} {Nat. Phys.}\ }\textbf {\bibinfo
  {volume} {13}},\ \bibinfo {pages} {545} (\bibinfo {year} {2017})}\BibitemShut
  {NoStop}%
\bibitem [{\citenamefont {Blanco~de Paz}\ \emph {et~al.}(2020)\citenamefont
  {Blanco~de Paz}, \citenamefont {Devescovi}, \citenamefont {Giedke},
  \citenamefont {Saenz}, \citenamefont {Vergniory}, \citenamefont {Bradlyn},
  \citenamefont {Bercioux},\ and\ \citenamefont
  {García‐Etxarri}}]{blanco_de_paz_tutorial_2020}%
  \BibitemOpen
  \bibfield  {author} {\bibinfo {author} {\bibfnamefont {M.}~\bibnamefont
  {Blanco~de Paz}}, \bibinfo {author} {\bibfnamefont {C.}~\bibnamefont
  {Devescovi}}, \bibinfo {author} {\bibfnamefont {G.}~\bibnamefont {Giedke}},
  \bibinfo {author} {\bibfnamefont {J.~J.}\ \bibnamefont {Saenz}}, \bibinfo
  {author} {\bibfnamefont {M.~G.}\ \bibnamefont {Vergniory}}, \bibinfo {author}
  {\bibfnamefont {B.}~\bibnamefont {Bradlyn}}, \bibinfo {author} {\bibfnamefont
  {D.}~\bibnamefont {Bercioux}},\ and\ \bibinfo {author} {\bibfnamefont
  {A.}~\bibnamefont {García‐Etxarri}},\ }\bibfield  {title} {\bibinfo
  {title} {Tutorial: {Computing} {Topological} {Invariants} in {2D} {Photonic}
  {Crystals}},\ }\href {https://doi.org/10.1002/qute.201900117} {\bibfield
  {journal} {\bibinfo  {journal} {Adv. Quantum Tech.}\ }\textbf {\bibinfo
  {volume} {3}},\ \bibinfo {pages} {1900117} (\bibinfo {year}
  {2020})}\BibitemShut {NoStop}%
\bibitem [{\citenamefont {\"Unal}\ \emph {et~al.}(2019)\citenamefont {\"Unal},
  \citenamefont {Seradjeh},\ and\ \citenamefont {Eckardt}}]{unal_how_2019}%
  \BibitemOpen
  \bibfield  {author} {\bibinfo {author} {\bibfnamefont {F.~N.}\ \bibnamefont
  {\"Unal}}, \bibinfo {author} {\bibfnamefont {B.}~\bibnamefont {Seradjeh}},\
  and\ \bibinfo {author} {\bibfnamefont {A.}~\bibnamefont {Eckardt}},\
  }\bibfield  {title} {\bibinfo {title} {How to {Directly} {Measure} {Floquet}
  {Topological} {Invariants} in {Optical} {Lattices}},\ }\href
  {https://doi.org/10.1103/PhysRevLett.122.253601} {\bibfield  {journal}
  {\bibinfo  {journal} {Phys. Rev. Lett.}\ }\textbf {\bibinfo {volume} {122}},\
  \bibinfo {pages} {253601} (\bibinfo {year} {2019})}\BibitemShut {NoStop}%
\bibitem [{\citenamefont {Delplace}\ \emph {et~al.}(2017)\citenamefont
  {Delplace}, \citenamefont {Fruchart},\ and\ \citenamefont
  {Tauber}}]{Delplace_Fruchart_Tauber}%
  \BibitemOpen
  \bibfield  {author} {\bibinfo {author} {\bibfnamefont {P.}~\bibnamefont
  {Delplace}}, \bibinfo {author} {\bibfnamefont {M.}~\bibnamefont {Fruchart}},\
  and\ \bibinfo {author} {\bibfnamefont {C.}~\bibnamefont {Tauber}},\
  }\bibfield  {title} {\bibinfo {title} {Phase rotation symmetry and the
  topology of oriented scattering networks},\ }\href
  {https://doi.org/10.1103/PhysRevB.95.205413} {\bibfield  {journal} {\bibinfo
  {journal} {Phys. Rev. B}\ }\textbf {\bibinfo {volume} {95}},\ \bibinfo
  {pages} {205413} (\bibinfo {year} {2017})}\BibitemShut {NoStop}%
\bibitem [{\citenamefont {Delplace}(2020)}]{delplace_topological_2020}%
  \BibitemOpen
  \bibfield  {author} {\bibinfo {author} {\bibfnamefont {P.}~\bibnamefont
  {Delplace}},\ }\bibfield  {title} {\bibinfo {title} {Topological chiral modes
  in random scattering networks},\ }\href
  {https://doi.org/10.21468/SciPostPhys.8.5.081} {\bibfield  {journal}
  {\bibinfo  {journal} {SciPost Phys.}\ }\textbf {\bibinfo {volume} {8}},\
  \bibinfo {pages} {081} (\bibinfo {year} {2020})}\BibitemShut {NoStop}%
\bibitem [{\citenamefont {Ryu}\ \emph {et~al.}(2010)\citenamefont {Ryu},
  \citenamefont {Schnyder}, \citenamefont {Furusaki},\ and\ \citenamefont
  {Ludwig}}]{Ryu_2010}%
  \BibitemOpen
  \bibfield  {author} {\bibinfo {author} {\bibfnamefont {S.}~\bibnamefont
  {Ryu}}, \bibinfo {author} {\bibfnamefont {A.~P.}\ \bibnamefont {Schnyder}},
  \bibinfo {author} {\bibfnamefont {A.}~\bibnamefont {Furusaki}},\ and\
  \bibinfo {author} {\bibfnamefont {A.~W.~W.}\ \bibnamefont {Ludwig}},\
  }\bibfield  {title} {\bibinfo {title} {Topological insulators and
  superconductors: tenfold way and dimensional hierarchy},\ }\href
  {https://doi.org/10.1088/1367-2630/12/6/065010} {\bibfield  {journal}
  {\bibinfo  {journal} {New J. Phys.}\ }\textbf {\bibinfo {volume} {12}},\
  \bibinfo {pages} {065010} (\bibinfo {year} {2010})}\BibitemShut {NoStop}%
\bibitem [{\citenamefont {Sato}\ and\ \citenamefont {Ando}(2017)}]{Sato2017}%
  \BibitemOpen
  \bibfield  {author} {\bibinfo {author} {\bibfnamefont {M.}~\bibnamefont
  {Sato}}\ and\ \bibinfo {author} {\bibfnamefont {Y.}~\bibnamefont {Ando}},\
  }\bibfield  {title} {\bibinfo {title} {Topological superconductors: a
  review},\ }\href {https://doi.org/10.1088/1361-6633/aa6ac7} {\bibfield
  {journal} {\bibinfo  {journal} {Reports on Progress in Physics}\ }\textbf
  {\bibinfo {volume} {80}},\ \bibinfo {pages} {076501} (\bibinfo {year}
  {2017})}\BibitemShut {NoStop}%
\bibitem [{\citenamefont {Sticlet}\ \emph {et~al.}(2012)\citenamefont
  {Sticlet}, \citenamefont {Pi\'echon}, \citenamefont {Fuchs}, \citenamefont
  {Kalugin},\ and\ \citenamefont {Simon}}]{Sticlet2012}%
  \BibitemOpen
  \bibfield  {author} {\bibinfo {author} {\bibfnamefont {D.}~\bibnamefont
  {Sticlet}}, \bibinfo {author} {\bibfnamefont {F.}~\bibnamefont {Pi\'echon}},
  \bibinfo {author} {\bibfnamefont {J.-N.}\ \bibnamefont {Fuchs}}, \bibinfo
  {author} {\bibfnamefont {P.}~\bibnamefont {Kalugin}},\ and\ \bibinfo {author}
  {\bibfnamefont {P.}~\bibnamefont {Simon}},\ }\bibfield  {title} {\bibinfo
  {title} {Geometrical engineering of a two-band chern insulator in two
  dimensions with arbitrary topological index},\ }\href
  {https://doi.org/10.1103/PhysRevB.85.165456} {\bibfield  {journal} {\bibinfo
  {journal} {Phys. Rev. B}\ }\textbf {\bibinfo {volume} {85}},\ \bibinfo
  {pages} {165456} (\bibinfo {year} {2012})}\BibitemShut {NoStop}%
\bibitem [{\citenamefont {Quandt}(2021)}]{Quandt2021}%
  \BibitemOpen
  \bibfield  {author} {\bibinfo {author} {\bibfnamefont {A.}~\bibnamefont
  {Quandt}},\ }\bibfield  {title} {\bibinfo {title} {Winding numbers,
  discriminants and topological phase transitions},\ }\href
  {https://doi.org/https://doi.org/10.1016/j.physb.2021.412867} {\bibfield
  {journal} {\bibinfo  {journal} {Physica B Condens. Matter}\ }\textbf
  {\bibinfo {volume} {612}},\ \bibinfo {pages} {412867} (\bibinfo {year}
  {2021})}\BibitemShut {NoStop}%
\bibitem [{\citenamefont {Delplace}\ \emph {et~al.}(2011)\citenamefont
  {Delplace}, \citenamefont {Ullmo},\ and\ \citenamefont
  {Montambaux}}]{Delplace2011}%
  \BibitemOpen
  \bibfield  {author} {\bibinfo {author} {\bibfnamefont {P.}~\bibnamefont
  {Delplace}}, \bibinfo {author} {\bibfnamefont {D.}~\bibnamefont {Ullmo}},\
  and\ \bibinfo {author} {\bibfnamefont {G.}~\bibnamefont {Montambaux}},\
  }\bibfield  {title} {\bibinfo {title} {Zak phase and the existence of edge
  states in graphene},\ }\href {https://doi.org/10.1103/PhysRevB.84.195452}
  {\bibfield  {journal} {\bibinfo  {journal} {Phys. Rev. B}\ }\textbf {\bibinfo
  {volume} {84}},\ \bibinfo {pages} {195452} (\bibinfo {year}
  {2011})}\BibitemShut {NoStop}%
\bibitem [{\citenamefont {P\'erez-Gonz\'alez}\ \emph
  {et~al.}(2019)\citenamefont {P\'erez-Gonz\'alez}, \citenamefont {Bello},
  \citenamefont {G\'omez-Le\'on},\ and\ \citenamefont
  {Platero}}]{perez-gonzalez_interplay_2019}%
  \BibitemOpen
  \bibfield  {author} {\bibinfo {author} {\bibfnamefont {B.}~\bibnamefont
  {P\'erez-Gonz\'alez}}, \bibinfo {author} {\bibfnamefont {M.}~\bibnamefont
  {Bello}}, \bibinfo {author} {\bibfnamefont {A.}~\bibnamefont
  {G\'omez-Le\'on}},\ and\ \bibinfo {author} {\bibfnamefont {G.}~\bibnamefont
  {Platero}},\ }\bibfield  {title} {\bibinfo {title} {Interplay between
  long-range hopping and disorder in topological systems},\ }\href
  {https://doi.org/10.1103/PhysRevB.99.035146} {\bibfield  {journal} {\bibinfo
  {journal} {Phys. Rev. B}\ }\textbf {\bibinfo {volume} {99}},\ \bibinfo
  {pages} {035146} (\bibinfo {year} {2019})}\BibitemShut {NoStop}%
\bibitem [{\citenamefont {Bernevig}\ and\ \citenamefont
  {Hughes}(2013)}]{Bernevig2013}%
  \BibitemOpen
  \bibfield  {author} {\bibinfo {author} {\bibfnamefont {B.~A.}\ \bibnamefont
  {Bernevig}}\ and\ \bibinfo {author} {\bibfnamefont {T.~L.}\ \bibnamefont
  {Hughes}},\ }\href@noop {} {\emph {\bibinfo {title} {Topological {Insulators}
  and {Topological} {Superconductors}}}}\ (\bibinfo  {publisher} {Princeton
  University Press},\ \bibinfo {year} {2013})\BibitemShut {NoStop}%
\bibitem [{\citenamefont {Jezequel}\ \emph {et~al.}(2022)\citenamefont
  {Jezequel}, \citenamefont {Tauber},\ and\ \citenamefont
  {Delplace}}]{Jezequel22}%
  \BibitemOpen
  \bibfield  {author} {\bibinfo {author} {\bibfnamefont {L.}~\bibnamefont
  {Jezequel}}, \bibinfo {author} {\bibfnamefont {C.}~\bibnamefont {Tauber}},\
  and\ \bibinfo {author} {\bibfnamefont {P.}~\bibnamefont {Delplace}},\
  }\bibfield  {title} {\bibinfo {title} {{Estimating bulk and edge topological
  indices in finite open chiral chains}},\ }\href
  {https://doi.org/10.1063/5.0096720} {\bibfield  {journal} {\bibinfo
  {journal} {J. Math. Phys.}\ }\textbf {\bibinfo {volume} {63}},\ \bibinfo
  {pages} {121901} (\bibinfo {year} {2022})}\BibitemShut {NoStop}%
\bibitem [{\citenamefont {Bessho}\ and\ \citenamefont
  {Sato}(2021)}]{Bessho2021}%
  \BibitemOpen
  \bibfield  {author} {\bibinfo {author} {\bibfnamefont {T.}~\bibnamefont
  {Bessho}}\ and\ \bibinfo {author} {\bibfnamefont {M.}~\bibnamefont {Sato}},\
  }\bibfield  {title} {\bibinfo {title} {Nielsen-ninomiya theorem with bulk
  topology: Duality in floquet and non-hermitian systems},\ }\href
  {https://doi.org/10.1103/PhysRevLett.127.196404} {\bibfield  {journal}
  {\bibinfo  {journal} {Phys. Rev. Lett.}\ }\textbf {\bibinfo {volume} {127}},\
  \bibinfo {pages} {196404} (\bibinfo {year} {2021})}\BibitemShut {NoStop}%
\bibitem [{\citenamefont {Regensburger}\ \emph {et~al.}(2011)\citenamefont
  {Regensburger}, \citenamefont {Bersch}, \citenamefont {Hinrichs},
  \citenamefont {Onishchukov}, \citenamefont {Schreiber}, \citenamefont
  {Silberhorn},\ and\ \citenamefont {Peschel}}]{Regensburger2011}%
  \BibitemOpen
  \bibfield  {author} {\bibinfo {author} {\bibfnamefont {A.}~\bibnamefont
  {Regensburger}}, \bibinfo {author} {\bibfnamefont {C.}~\bibnamefont
  {Bersch}}, \bibinfo {author} {\bibfnamefont {B.}~\bibnamefont {Hinrichs}},
  \bibinfo {author} {\bibfnamefont {G.}~\bibnamefont {Onishchukov}}, \bibinfo
  {author} {\bibfnamefont {A.}~\bibnamefont {Schreiber}}, \bibinfo {author}
  {\bibfnamefont {C.}~\bibnamefont {Silberhorn}},\ and\ \bibinfo {author}
  {\bibfnamefont {U.}~\bibnamefont {Peschel}},\ }\bibfield  {title} {\bibinfo
  {title} {Photon {Propagation} in a {Discrete} {Fiber} {Network}: {An}
  {Interplay} of {Coherence} and {Losses}},\ }\href
  {https://doi.org/10.1103/PhysRevLett.107.233902} {\bibfield  {journal}
  {\bibinfo  {journal} {Phys. Rev. Lett.}\ }\textbf {\bibinfo {volume} {107}},\
  \bibinfo {pages} {233902} (\bibinfo {year} {2011})}\BibitemShut {NoStop}%
\end{thebibliography}%
\clearpage
\newpage
\appendix

\section{Data processing \label{app:data_processing}}
In this section, we will explain how to extract the eigenvectors, construct the Berry curvature and calculate the Chern number from the experimental data.

The lattice has two dimensions: a real-space one, defined by the sites $n$, and a parametric one whose coordinates are given by the value $\varphi$ of the phase modulator in one of the rings. To reconstruct the two-dimensional dispersion  and associated eigenvectors we extract the eigenvalues and eigenvectors from the measured dynamics for a given value of $\varphi$, which is varied from $-\pi$ to $\pi$ in successive experiments. For each experiment, we use an arbitrary waveform generator (AWG) to generate two consecutive 1.4~ns pulses (a calibration pulse and a science pulse) separated by a time delay of 90-round trips  $T_R$.
For the calibration pulse the beamsplitter is set to $\theta=\pi/4$, which corresponds to a 50/50 splitting ratio, and $\varphi=0$. The time dynamics of this model is quite trivial and well known~\cite{Regensburger2011}, and serves as a reference model. The science shot describes the two-steps model we wish to study, in which the variable beam splitter alternates between two coupling values $\theta_1$ and $\theta_2$, and the external phase modulator adds a controlled phase $\varphi$ with a value that alternates between $\varphi$ and $-\varphi$ in odd and even steps, respectively.

\begin{figure}[b]
\includegraphics[width=\columnwidth]{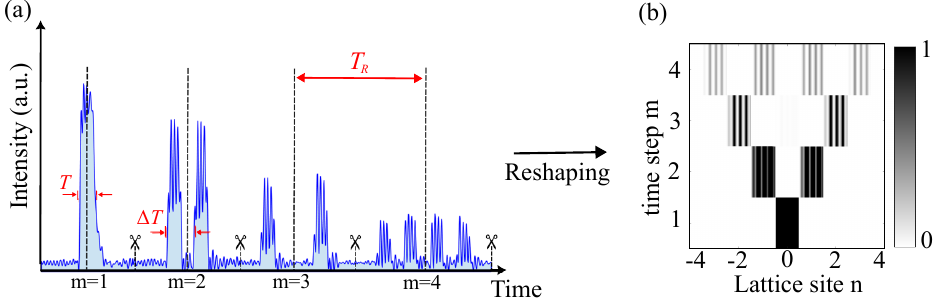}
\caption{\label{fig:supp_pulses} 
    (a) Zoom on the first time steps of the measured time trace of the signal intensity at the output of the $\alpha$ fiber loop. (b) Space-time diagram of the $\alpha$ ring reconstructed from the time trace in (a).
    }
\end{figure}

For each injected pulse, we record on the oscilloscope the signal intensity at the output of both fiber loops as a function of time. After the injected calibration or science pulse we observe a series of groups of pulses separated by a time  $T_R =225$~ns. This time corresponds to the average round trip time inside the loops. Each pulse has a temporal width of $T=1.4$~ns and is separated from adjacent pulses in the same group a time interval $\Delta$T =2.8 ns, corresponding to the length difference between the two fiber loops. To clearly observe separate sequences of pulses and avoid any overlap, the width of each pulse must be shorter than the time difference separating the two rings ($\Delta$T $<$  T). Figure~\ref{fig:supp_pulses}(a) shows a typical measured time trace of the ~$\beta$ ring for the first roundtrips after the arrival of the calibration pulse.

%%%%%%%%%%%%%%%%%%%%%%%%%%%%%%%%%%%%%%%%%%%%%%%%%%%%%
%%%%%%%%%%%%%%%%%%%%%%%%%%%%%%%%%%%%%%%%%%%%%%%%%%%%%

To create the space-time diagram of the evolution of light intensity as a function of position and time in the $\alpha$ and $\beta$ rings, we cut each recorded time trace after the arrival of the injected pulse into equal time segments of duration $T_R$ (see Fig.~\ref{fig:supp_pulses}(a)). The segments are then arranged in lines corresponding to each round trip time $m$, the spatial position $n$ denoting the relative time of each pulse within each segment (see Fig.\ref{fig:supp_pulses}(b)). This procedure can be applied to either the first part of the measured trace, corresponding to the dynamics after the injection of the calibration pulse, or to the second part of the time trace, corresponding to the dynamics after the injection of the science pulse. Note that in between the two pulses, high losses are applied to both rings for a few miliseconds using an electronic switch to make sure that no light from the calibration experiment is present during the science experiment. The reason to do the two experiments in the same oscilloscope shot will become clear later.

\begin{figure}[t]
\includegraphics[width=\columnwidth]{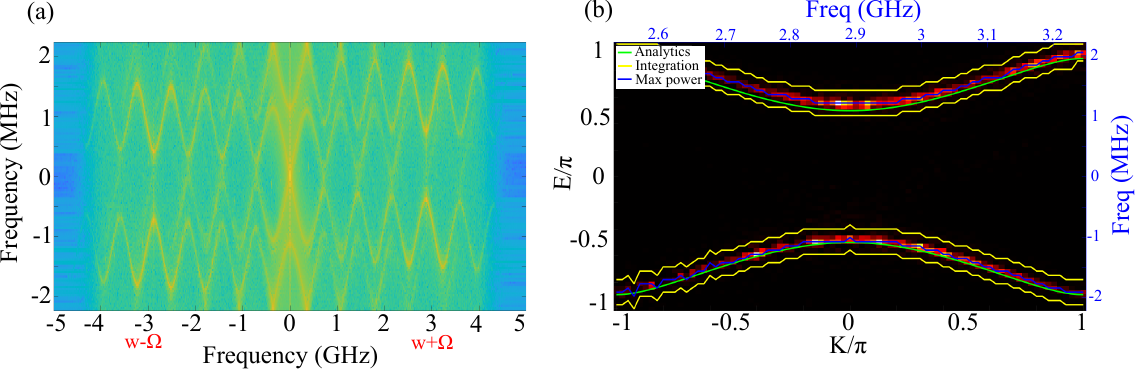}
\caption{\label{fig:suppl_bands} 
    (a) Two-dimensional Fourier transform of the measured space-time dynamics of the calibration part of the time trace. (b) Measured bands in one Brillouin zone.
    }
\end{figure}

To measure the excited eigenvectors and the corresponding eigenvalues we need information about the relative phase of the pulses in the rings at different spatial sites $n$ and time steps $m$. Because the photodiodes are only sensitive to intensity, we reveal the relative phase by interfering the signal that exits the rings at each round trip with a reference continuous wave laser. This reference laser is shifted in frequency by  about 3~GHz with respect to the laser that creates the injected pulses and has a coherence length of hundreds of  microseconds, much longer than the duration of each recorded time trace. The interference fringes of the two lasers are visible in Fig.~\ref{fig:supp_pulses}. By doing a two-dimensional Fourier transform of the space-time diagram shown in Fig.~\ref{fig:supp_pulses}(b) and in Figs.~\ref{fig:spatiotemporal}(a) and~(b) we obtain the diagram shown in Fig.~\ref{fig:suppl_bands}(a)). This figure shows the periodic eigenvalue bands spanning in the range of 10 GHz in the horizontal direction and 4~MHz in the vertical axis  (see Fig.~\ref{fig:suppl_bands}(a)). These frequencies correspond to the conjugate times within each segment, and from time step to time step. Therefore, they have the meaning of quasimomentum and quasienergy.

%%%%%%%%%%%%%%%%%%%%%%%%%%%%%%%%%%%%%%%%%%%%%%%%%%%%%5
%%%%%%%%%%%%%%%%%%%%%%%%%%%%%%%%%%%%%%%%%%%%%%%%%%%%%5

We focus on a single Brillouin zone (Fig.~\ref{fig:suppl_bands}(b)) at around a frequency of 3~GHz. From shot to shot, the Fourier transformed diagram moves slightly due to variations in laser wavelength and fibre length. To fine-center the selected zone, we use a correlator between the measured data and the analytic dispersion of the model in the calibration shot. Note that once this is done for the calibration shot, the Brillouin zone window remains exactly the same for the science dynamics. This is one of the reasons to design the experiment with this double injection protocol in the same experimental shot.

To identify the bands of eigenvalues from the Fourier transform of the measured spatiotemporal  diagrams we proceed in the following way. For each value of quasimomentum, we use the analytically computed band corresponding to the nominal value of the external phase modulator as a reference (green line Fig.~\ref{fig:suppl_bands}(b)). To locate the maximum of intensity of the experimental bands (blue line Fig.~\ref{fig:suppl_bands}(b)) we scan two pixels in quasienergy above and below the analytical bands (yellow line Fig.~\ref{fig:suppl_bands}(b)). Other more general methods based exclusively on the location of the maxima of intensity extracted from the experiment are also successful.

To extract the eigenvectors, we take the complex amplitude of the measured bands at each point in quasimomentum  for the $\alpha$ and $\beta$ rings. To reduce the noise in the masurement of the ratio of amplitudes $R$, we integrate the recorded intensity over two pixels below and above the maximum of the band for each value of $k$. The ratio $R$ and the phase difference $\Phi_{\alpha \beta}$ of the complex amplitudes directly provide the eigenvectors Eq.~(\ref{eigenmodes}).

For the calibration shot, the relative phase difference is well defined for all values of quasimomenta $k$  up to an overall rigid shift $\Delta \Phi$. From shot to shot, this shift may vary due to slight differences in length in the fibre components of the experiment. For this reason, in each measured shot, we rigidly shift $\Phi_{\alpha \beta}(k=-\pi)$ to zero in the calibration shot. When we treat the science part of the time trace, we follow the exact same procedure to extract the eigenvectors, and we apply the rigid shift $\Delta \Phi$ found in the calibration trace of the same shot. In this way we are sure of having an unambiguous phase reference for the measured  $\Phi_{\alpha \beta}(k)$ in the science bands.

Following this procedure is crucial when reconstructing the two-dimensional sublattice phase pattern $\Phi_{\alpha \beta}(k, \varphi)$ shown in Fig.~\ref{fig2}. As already mentioned, these diagrams are constructed from independent measurements for different values of $\varphi \in [-\pi, \pi]$. For the values of $\Phi_{\alpha \beta}(k, \varphi)$ to be consistent when changing $\varphi$, we need the phase reference of the calibration shot.

Finally, we perform a number of averaging procedures. First, the measured ratio of amplitude $R$ and phase $\Phi_{\alpha \beta}$ of the eigenvectors are averaged over five to ten identical experiments for each value of $\varphi$. Then, we apply a Gaussian averaging ("smoothdata" function in Matlab) to the two-dimensional matrices $R(k, \varphi)$ and $\Phi_{\alpha \beta}(k, \varphi)$. The result is displayed in Fig.~\ref{fig2} for the upper band.

\begin{figure}[t!]
\centering
\includegraphics[width=0.7\columnwidth]{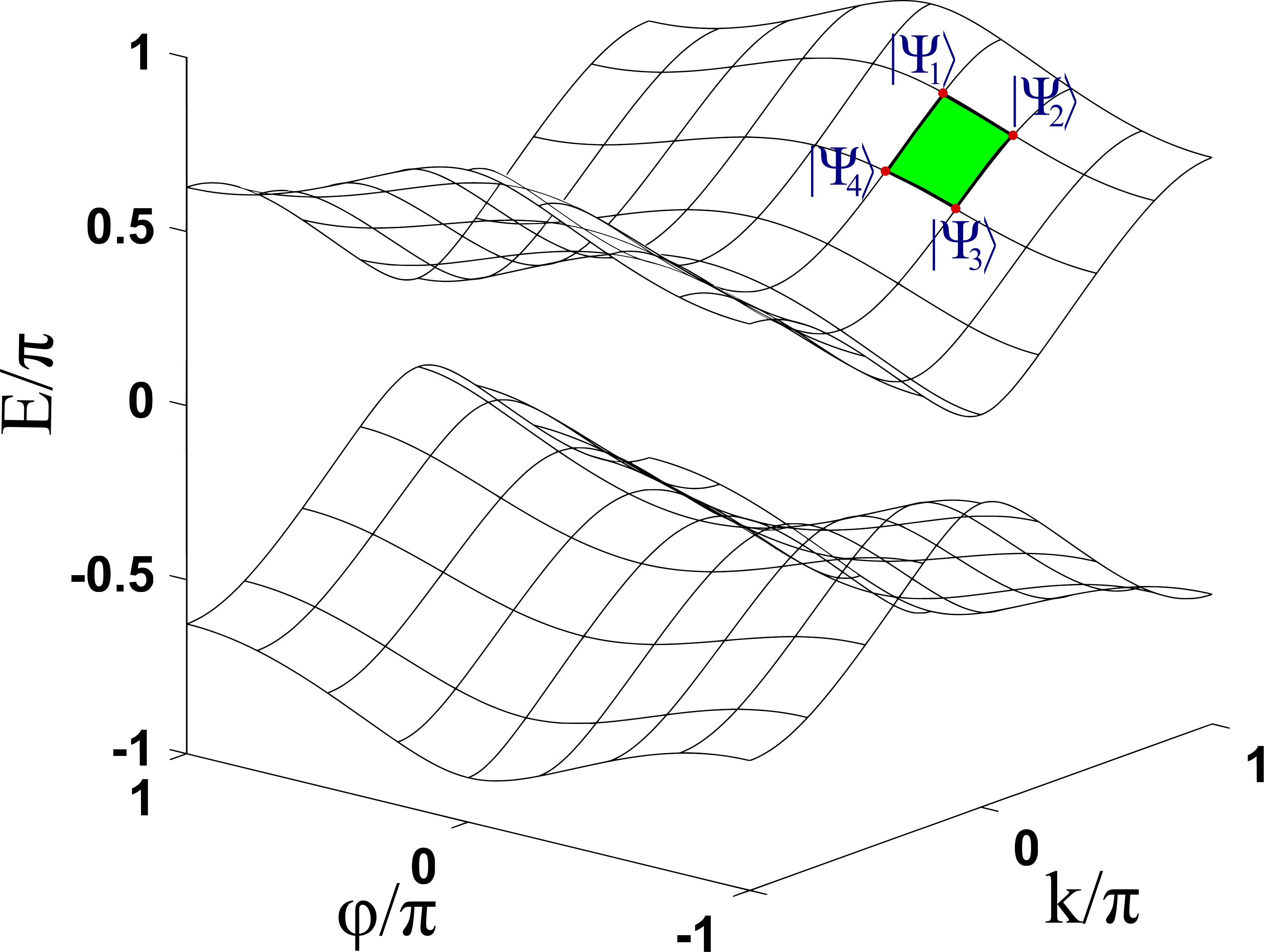}
\caption{\label{fig:suppl_Berry} 
    Discretization of the bands in two dimensions over the Brillouin zone.
    }
\end{figure}

\begin{figure*}[t]
\includegraphics[width=\textwidth]{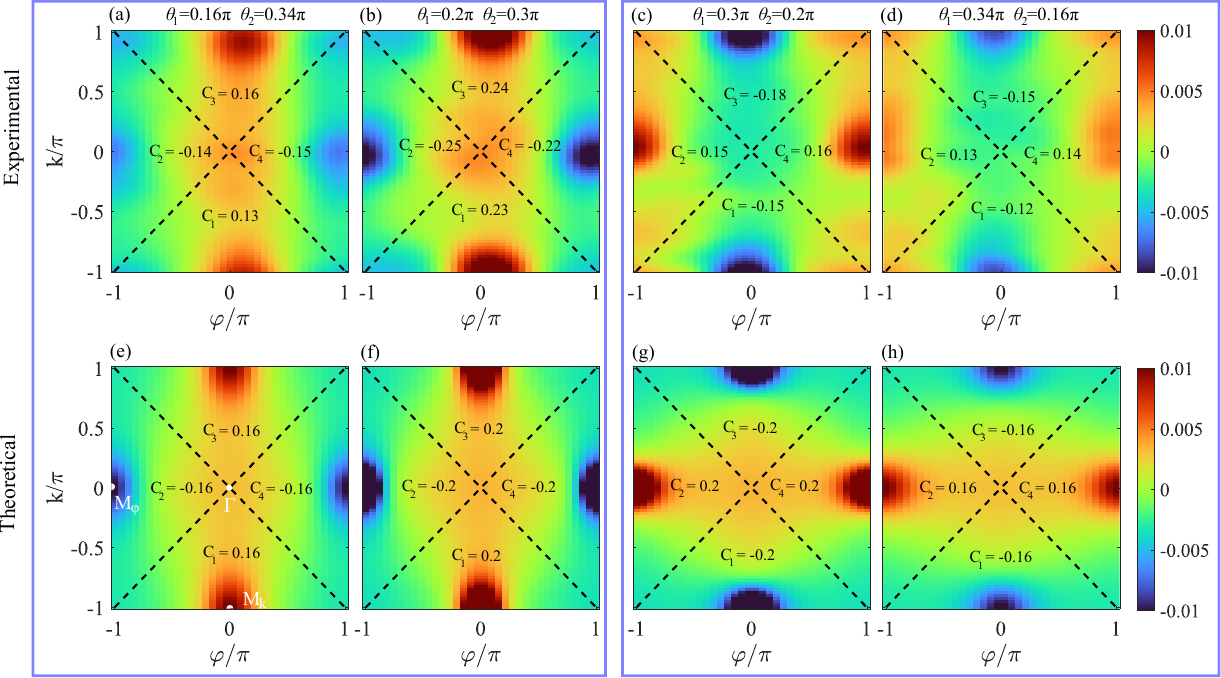}
\caption{\label{fig:suppl_Berry_lower} 
    Measured (upper panels) and analytically computed Berry curvature for the lower band at points in the phase diagram displayed in Fig.~\ref{fig2}(f). The numbers in each quadrant show the integrated Berry curvature in the quadrant.
    }
\end{figure*}

To compute the Berry curvature we use the natural discretization of the Brillouin zone coming from the experimental data. The Berry curvature can be computed by taking the product of the eigenvectors of a given band at the four corners of the square of the discretized Brillouin zone (see Fig.~\ref{fig:suppl_Berry}):
\begin{equation}
\text{BC}= - \text{Im}\: \log \: [\langle\psi_{1}|\psi_{2}\rangle\langle\psi_{2}|\psi_3\rangle\langle\psi_3|\psi_4\rangle\langle\psi_4|\psi_1\rangle].
\end{equation}
Figure~\ref{fig2} shows the measured and analytically calculated Berry curvature for the upper band at different points in the phase diagram. 
Figure~\ref{fig:suppl_Berry_lower} shows the same for the lower band.
Finally, the Chern number is computed by integrating the Berry curvature over the entire Brillouin zone.

\begin{figure*}[t]
    \includegraphics[width=\textwidth]{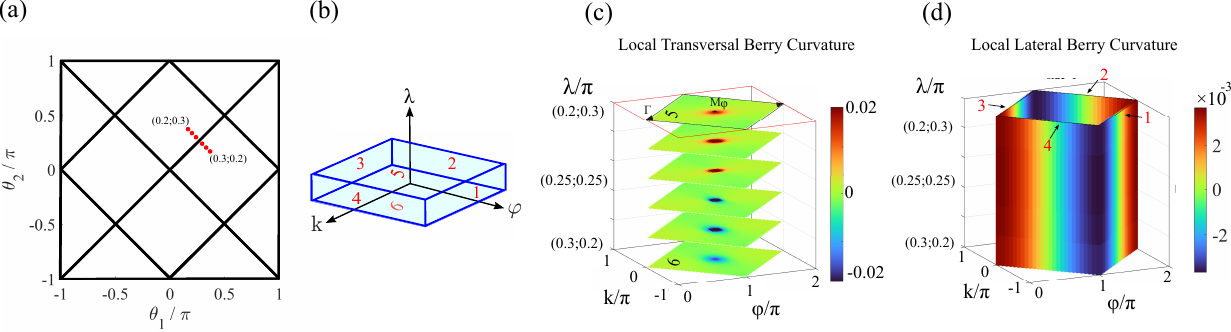}
    \caption{(a) Topological phase diagram of the system as a function of the splitting angles $\theta_1$ and $\theta_2$. (b) Cuboid in parameter space where $\lambda$ labels the evolution of the splitting amplitudes $\theta_1$ and $\theta_2$ across the phase transition. Faces 1-4 denote the local Berry flux through the lateral faces, while faces 5 and 6 represent the transversal local Berry flux over a region of the standard Brillouin zone. (c) Analytical computation of the transversal Berry flux when varying the parameters $\theta_1$,$\theta_2$ across the phase transition from the upper side (5) to the lower side (6) of the cuboid. (d) Computed Berry curvature through the lateral faces 1-4 of the cuboid when $(\theta_1$,$\theta_2)$ evolve from (0.2, 0.3)$\pi$ to (0.3, 0.2)$\pi$. The integrated Berry flux through each of the lateral faces is zero.} 

    \label{fig15}
\end{figure*}

\begin{figure*}[t]
    \includegraphics[width=\textwidth]{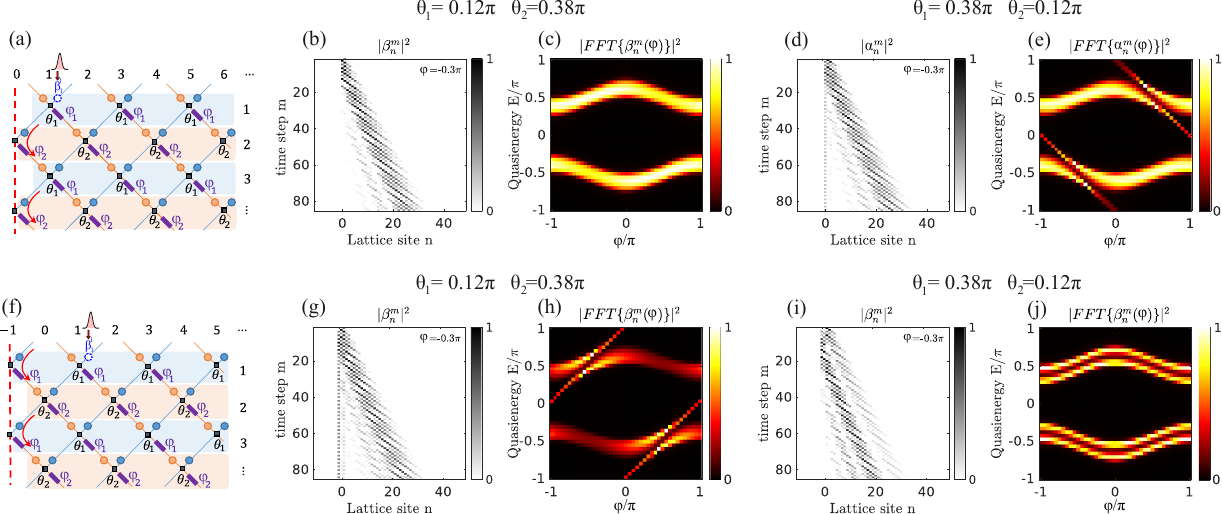}
    \caption{Numerical simulation of the spatiotemporal dynamics and the associated bands in the exact same conditions as in Fig.~\ref{fig:edge}.}
    \label{fig:edge_sim}
\end{figure*}

For completeness, Fig.~\ref{fig15} displays the computed Berry curvature through the sides of the cuboid represented in Fig.~\ref{fig2}(g). Figure~\ref{fig:edge_sim} shows numerical simulations of the spatiotemporal dynamics and the associated bands in the exact same conditions as in Fig.~\ref{fig:edge}.

%%%%%%%%%%%%%%%%%%%%%%%%%%%%%%%%%%%%%%%%%%%%%%%%%%%%%5
%%%%%%%%%%%%%%%%%%%%%%%%%%%%%%%%%%%%%%%%%%%%%%%%%%%%%5

%%%%%%%%%%%%%%%%%%%%%%%%%%%%%%%%%%%%%%%%%%%%%%%%%%%%%5
%%%%%%%%%%%%%%%%%%%%%%%%%%%%%%%%%%%%%%%%%%%%%%%%%%%%%5

\section{Calculation of the eigenvectors and eignenvalues}
The eigenvectors of the evolution Eqs.~(\ref{Eq:step}) are represented by the complex amplitudes $\alpha$ and $\beta$ of each of the two sublattices, Eq.~(\ref{eigenmodes}). The analytic value of $\lvert R \rvert$ and, therefore, of the eigenvectors, can be straighforwardly calculated by applying the Floquet-Bloch ansatz to the two time steps evolution Eqs.~(\ref{Eq:C4})-(\ref{Eq:C5}). The solution is:
\begin{flalign}
\label{eigenvector 1}
R(k,\varphi) &= \lvert R \rvert e^{i\Phi_{\alpha\beta}} \notag \\
&=\frac {[ e^{iE(k,\varphi)} -\cos\theta_2\cos\theta_1  e^{- i k}  +\sin\theta_2\sin\theta_1 e^{-i\varphi} ]}{[i \cos\theta_2\sin\theta_1 e^{-i k}  +i\sin\theta_2\cos\theta_1  e^{-i\varphi}]}
\end{flalign}
In the above expression, the eigenvalues $E(k,\varphi)$ take different magnitudes for each of the bands $\pm$:
\begin{flalign}
    &E_{\pm}(k,\varphi)=
    \pm\cos^{-1}[\cos\theta_2\cos\theta_1 \cos(k)
    -\sin\theta_2\sin\theta_1\cos(\varphi) ]  &&
\end{flalign}

\begin{figure*}[t!]
    \includegraphics[width=\textwidth]{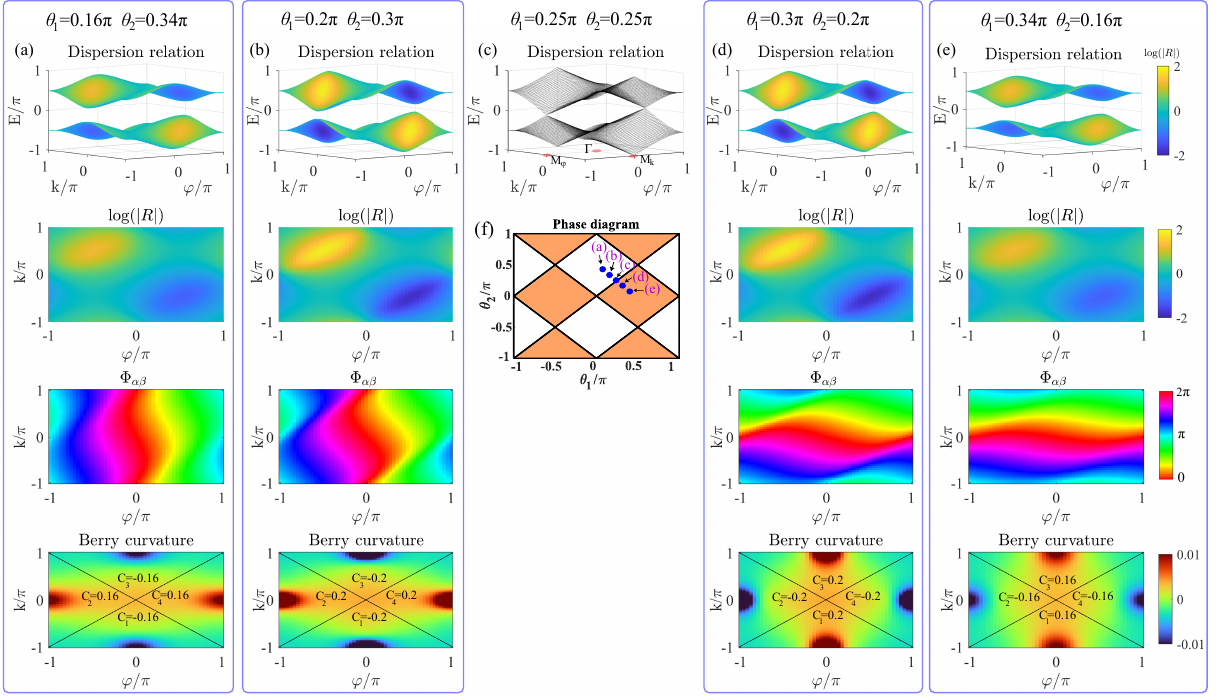}
    \caption{Analytical eigenvalues (top row), and the relative amplitude and phase difference between the two sublattices for the eigenvectors corresponding to the upper band in the conditions of Fig.~\ref{fig2}. The bottom row displays the Berry curvature computed from the analytic eigenvalues. \label{fig16}}
\end{figure*}

Figure~\ref{fig16} displays the analytical eigenvalue bands and the relative amplitude and phase difference between the two sublattices for the eigenmodes of the upper band in the conditions of Fig.~\ref{fig2}, in which most of the experiments were realized.

%%%%%%%%%%%%%%%%%%%%%%%%%%%%%%%%%%%

%%%%%%%%%%%%%%%%%%%%%%%%%%%%%%%%%%%%%%%%%%%%%%%%%%%%%

%%%%%%%%%%%%%%%%%%%%%%%%%%%%%%%%%%%
%%%%%%%%%%%%%%%%%%%%%%%%%%%%%%%%%%%
%%%%%%%%%%%%%%%%%%%%%%%%%%%%%%%%%%%
%%%%%%%%%%%%%%%%%%%%%%%%%%%%%%%%%%%

%%%%%%%%%%%%%%%%%%%%%%%%%%%%%%%%%%%%%%%%%%%%%%%%%%%%%

%%%%%%%%%%%%%%%%%%%%%%%%%%%%%%%%%%%

\section{Topological characterization\label{appendix:Invariant}}
In this Appendix we discuss the topological characterization of the system and present the detailed calculation of the invariant for the quantum walk.\\
The general time evolution of the initial state for our 2-step protocol can be written as the product of two different unitaries:
\begin{equation}
    U=\ldots U_2 U_1 U_2 U_1 U_2 U_1 \ldots
\end{equation}
Therefore, the stroboscopic time-evolution over a period or Floquet operator for this protocol corresponds to the product of two unitaries: $U_\text{QW} = U_2 U_1$, and as it propagates the system over many periods, we can also change the order to $\tilde{U}_\text{QW} = U_1 U_2$, and represent an identical protocol.
This description only shifts the reference frame one time step, and can be checked that the corresponding quasienergies are identical for the two cases.\\
Let us now particularize to our protocol, which is described by the following equations of motion at each time-step:

\begin{align}
    \alpha_{n}^{m+1}	&=\alpha_{n-1}^{m}\cos\left(\theta_{m}\right)e^{i\varphi_{m}}+i\beta_{n-1}^{m}\sin\left(\theta_{m}\right)e^{i\varphi_{m}}\\
\beta_{n}^{m+1}	&=i\alpha_{n+1}^{m}\sin\left(\theta_{m}\right)+\beta_{n+1}^{m}\cos\left(\theta_{m}\right)
\end{align}
Then, for our 2-step protocol we can write the equations of motion describing the time evolution during each full period as:

\begin{flalign}
&\alpha_{n}^{m+2} =	\left[-\alpha_{n}^{m}\sin\left(\theta_{m}\right)+i\beta_{n}^{m}\cos\left(\theta_{m}\right)\right]\sin\left(\theta_{m+1}\right)e^{i\varphi_{m+1}}\notag\\
&+\left[\alpha_{n-2}^{m}\cos\left(\theta_{m}\right)+i\beta_{n-2}^{m}\sin\left(\theta_{m}\right)\right]\cos\left(\theta_{m+1}\right)e^{i(\varphi_{m}+\varphi_{m+1})}\label{Eq:C4}\\
&\beta_{n}^{m+2} =	\left[i\alpha_{n}^{m}\cos\left(\theta_{m}\right)-\beta_{n}^{m}\sin\left(\theta_{m}\right)\right]\sin\left(\theta_{m+1}\right)e^{i\varphi_{m}}\notag\\
&+\left[i\alpha_{n+2}^{m}\sin\left(\theta_{m}\right)+\beta_{n+2}^{m}\cos\left(\theta_{m}\right)\right]\cos\left(\theta_{m+1}\right)\label{Eq:C5}
\end{flalign}
Now, we fix our reference frame by assuming that $U_1$ acts before $U_2$ and produce the synthetic dimension by requiring $\varphi_{m}=-\varphi_{m+1}=\varphi$. Then, for periodic boundary conditions, the matrix describing one period of the quantum walk becomes:
\begin{equation}
    U_{\text{QW}}\left(\mathbf{q}\right)=	v_{0}\sigma_{0}+i\mathbf{v}\left(\mathbf{q}\right)\cdot\boldsymbol{\sigma}
\end{equation}
with $\mathbf{q}=(k,\varphi)$, the vector components:
\begin{align}
    v_{0}=&J_{1}\cos\left(k\right)-m_{1}\cos\left(\varphi\right)\\
    v_{x}=&J_{2}\cos\left(k\right)+m_{2}\cos\left(\varphi\right)\\
    v_{y}=&J_{2}\sin\left(k\right)+m_{2}\sin\left(\varphi\right)\\
    v_{z}=&-J_{1}\sin\left(k\right)+m_{1}\sin\left(\varphi\right)
\end{align}
and the parameters $J_{1} = \cos\left( \theta_{1} \right) \cos \left( \theta_{2} \right)$, $J_{2} = \sin\left(\theta_{1}\right) \cos\left(\theta_{2}\right)$, $m_{1} = \sin\left(\theta_{1}\right) \sin\left(\theta_{2}\right)$ and $m_{2} = \cos\left(\theta_{1}\right) \sin\left(\theta_{2}\right)$. In addition, unitarity imposes the constrains $J_{1}m_{1} = J_{2}m_{2}$ and $J_{1}^{2}+J_{2}^{2} + m_{1}^{2}+m_{2}^{2} =1$, because the four parameters are not independent (only $\theta_{1,2}$ are). Notice that the shifted Floquet operator, $\tilde{U}_\text{QW} = U_1 U_2$, can be easily obtained by just swapping $\theta_1\leftrightarrow \theta_2$ and $\varphi \leftrightarrow -\varphi$.\\
To study the bulk topology one must look at the symmetry class of $U_\text{QW}(\mathbf{q})$. It can be checked the presence of particle-hole symmetry (PHS):
\begin{equation}
    \mathcal{C}U_{\text{QW}}\left(\mathbf{q}\right)\mathcal{C}^{-1} = U_{\text{QW}}\left(-\mathbf{q}\right)
\end{equation}
for the anti-unitary operator $\mathcal{C}=\sigma_{z}K$, with $K$ the conjugation operation and $\sigma_z$ the third Pauli matrix. Also, the PHS operator fulfills $\mathcal{C}^{2}=+1$, placing the Floquet operator in D class in 2 dimensions, which is characterized by a Chern number.\\
%See Fig.~\ref{fig:Appendix1} for a plot of the Berry curvature distribution over the first Brillouin zone.
%\begin{figure}
%    \centering
%    \includegraphics[width=0.9\columnwidth]{Appendix1.pdf}
%    \caption{Berry curvature distribution over the FBZ for $\theta_{1,2}=(\pi/8,\pi/4)$. The sum over the whole FBZ results in a vanishing Chern number, $c$, for all parameters and the two unequivalent gaps.\alvaro{[plot from $-\pi$ to $\pi$ instead and fix legend scale.]}}
%    \label{fig:Appendix1}
%\end{figure}
The calculation of the Chern number shows that it vanishes for all values of $\theta_{1,2}$, indicating that the topology, if any, must be anomalous.\\
Importantly, the $0$- and $\pi$-gaps simultaneously close when $\theta_1\pm\theta_2=n\pi$ for $n\in\mathbb{Z}$. However, one should not expect a different bulk topology at either side of the gap closing point, because the bulk topology should not depend on the reference frame, and the two points are related by the swap $\theta_1 \leftrightarrow \theta_2$, which is equivalent to a shift of the time-frame of half a period. This is in agreement with the vanishing of the Chern number for all $\theta_{1,2}$.\\

\begin{figure}[b!]
    %\centering
    \includegraphics[width=0.49\columnwidth]{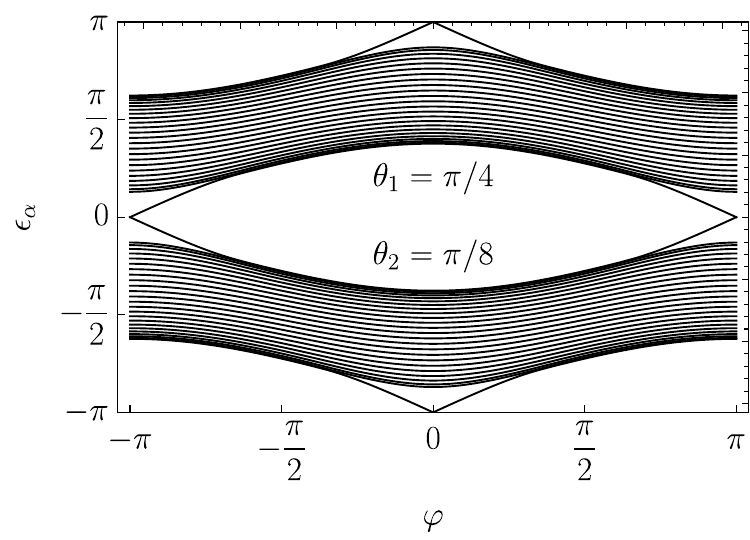}
    \includegraphics[width=0.49\columnwidth]{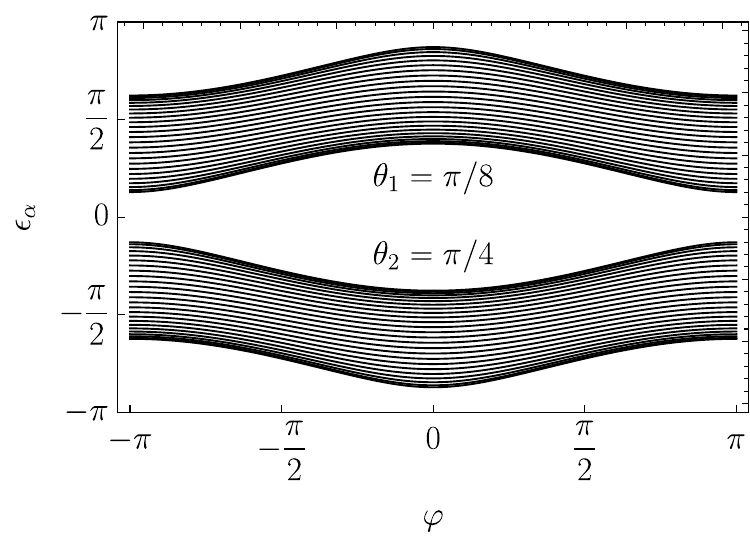}
    \caption{Spectra for a ribbon with PBC along the synthetic coordinate at symmetric values of $\theta_{1,2}$ before and after the gap closure. The bulk is identical, but the left plot describes an anomalous phase with chiral edge states in each gap, while the right plot lacks the edge states.}
    \label{fig:Appendix2}
\end{figure}

To study if the topology is trivial or anomalous, we first calculate the spectrum for a ribbon with OBC along the spatial coordinate, and periodic boundary conditions along the synthetic one (see Fig.~\ref{fig:edge}(a) for a schematic). 
This choice corresponds to the one of the experimental setup, where OBC are implemented as fully reflecting edges which can separate two dimers, or the two sites within a dimer. Notice that the two cases are also related by a reference frame shift of half a period.\\
The result is shown in Fig.~\ref{fig:Appendix2}. Surprisingly, it shows that for $\theta_1=\pi/4$ and $\theta_2=\pi/8$, chiral edge states populate both gaps, indicating the presence of an anomalous phase.
%%%%%%%%%%%%%%%%%%%%%%%%%%%%%%%%%%%%
%%%%%%%%%%%%%%%%%%%%%%%%%%%%%%%%%%%
%%%%%%%%%%%%%%%%%%%%%%%%%%%%%%%%%%%%
%%%%%%%%%%%%%%%%%%%%%%%%%%%%%%%%%%%%%%
In contrast, swapping the angles $\theta_1\leftrightarrow\theta_2$ the edge states disappear. This seems to be in contradiction with the fact that swapping the angles should not affect the topology, because it is equivalent to a reference frame time-shift.

There are two equivalent ways to reconcile these ideas, and both highlight the relevance of boundaries in the topological characterization of quantum walks: the first way invokes extrinsic topology~\cite{Bessho2022}, which explains that in quantum walks there is intrinsic topology, which is encoded in the bulk effective Hamiltonian of the Floquet operator $H_\text{F}(\mathbf{q})=i\log[U_{\text{QW}}(\mathbf{q})]$, but also extrinsic topology, which is produced when OBCs are implemented, due to the unitary nature of the boundary operators and their possibility to have non-vanishing winding number, unlike Hermitian operators.
Both contribute to the presence of edge states in the system, and in particular, extrinsic topology is the one responsible for anomalous phases, because the boundary unitary operators directly produce chiral edge modes along both gaps. That is, it plays a similar role as the micromotion in periodically driven Hamiltonians.\\
Alternatively, here we show that the role of boundaries and edge unitaries can be linked with different Floquet operators with PBC in different time-frames, and that their winding number at high-symmetry points completely characterizes the presence of edge states.
This interpretation implies that for the same parameters $\theta_{1,2}$ as in Fig.~\ref{fig:Appendix2}(right) we could obtain edge states by just changing the boundary conditions.
We have experimentally confirmed this, by considering a fully reflecting edge at a neighboring site and checking that the chiral edge states appear. This is shown in Fig.~\ref{fig:edge} of the main text.  

To find the topological invariant, let us first notice that PHS in topological systems in D class imprints tight restrictions on the trajectories of the Bloch vector~\cite{Sato2017}. In particular, at high-symmetry points, $\mathbf{q}=\pi\left(n,m\right)$ with $n,m=\left\{0,\pm 1\right\}$, the Bloch vector aligns with the $x$-axis, $\mathbf{v}=\left( v_x, 0, 0 \right)$.
As in our case the OBC is always along the spatial coordinate and the PBC along the synthetic dimension, $\varphi$, we are interested in topologically nonequivalent loops as $k$ is varied. 
To simplify the analysis, let us first focus on the high symmetry points along the synthetic dimension $\varphi=m\pi$.
In that case the Bloch vector actually describes a system with chiral symmetry
\begin{equation}
    \Gamma U_{\text{QW}}(k,m\pi) \Gamma^{-1} = U_{\text{QW}}(k,m\pi)^{\dagger}
\end{equation}
Furthermore, the chiral operator can be easily obtained as
\begin{equation}
    \Gamma\left(\varphi\right) = i e^{-i\pi\vec{n}_{\Gamma}\left(\varphi\right)\cdot\vec{\sigma}/2}
\end{equation}
for $\varphi=\left\{ 0,\pi\right\}$, and $\vec{n}_{\Gamma}\left(\varphi\right)$ the vector orthogonal to the plane of rotation.
In particular, at the high symmetry points it is given by: $\vec{n}_{\Gamma}\left(\varphi\right)=\left(0,J_{1},J_{2}\right)\left(J_{2}\pm m_{2}\right)$ for $\varphi=0$ and $\pi$, respectively.

\begin{figure*}[t!]
    \centering
    \includegraphics[width=1\textwidth]{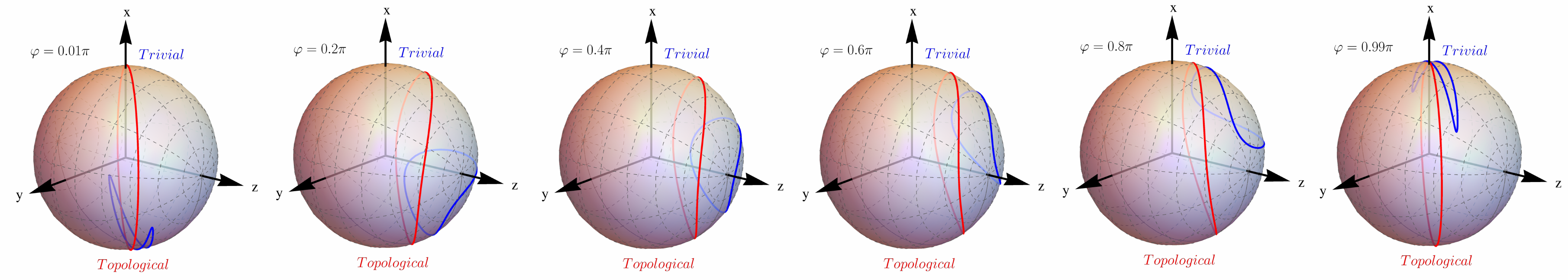}
    \caption{Topologically trivial (blue) and non-trivial (red) trajectories, for different values of the parameter $\varphi$. At high symmetry points $\varphi=0,\pi$, the trivial loop crosses twice the same pole, while the non-trivial one crosses each pole only once. Then, as $\varphi$ moves away from the high symmetry point, chiral symmetry breaks and trajectories are deformed out-of-plane. However, due to the presence of PHS, the two trajectories remain topologically nonequivalent for all values of $\varphi$.}
    \label{fig:CompositeTrajectories}
\end{figure*}

For this case, the components of the Bloch vector at the high symmetry points $\varphi=\left\{0,\pi\right\}$, read:
\begin{align}
    v_{x}=&J_{2}\cos\left(k\right)\pm m_{2},\\
    v_{y}=&J_{2}\sin\left(k\right),\\
    v_{z}=&-J_{1}\sin\left(k\right),
\end{align}
In 1D systems with chiral symmetry it is possible to calculate the winding number straightforwardly~\cite{Sticlet2012}:
\begin{align}
    \nu &= \frac{1}{2\pi}\int_{-\pi}^{\pi} dk \frac{\mathbf{v}\times \partial_k\mathbf{v}}{  v^2} \cdot \hat{\mathbf{n}}\\
    &= \sum_{\mathbf{v}(k_i)=\mathbf{v}_0} \text{sgn} [\mathbf{v}\times \partial_k\mathbf{v}) \cdot \hat{\mathbf{n}}]
\end{align}
However, this calculation is not even necessary in our case, because the only two types of possible trajectories at high symmetry points can be classified as those that pass through both poles of the Bloch sphere, or those that only return to the original pole without reaching the other (see Figs.~\ref{fig:trajectories} and~\ref{fig:CompositeTrajectories}).
This means that we only need to know if the vector passes through different poles at $k=\left\{0,\pi\right\}$ and we can write the invariant as:
\begin{align}
\label{inv_supp}
    \nu	=& \frac{1}{2}\left[ 1-\text{sgn}\left(J_{2}+m_{2}\right)\text{sgn}\left(m_{2}-J_{2}\right)\right] \nonumber \\
	=& \frac{1}{2}\left[ 1 - \text{sgn}\left[\sin\left(\theta_{1}+\theta_{2}\right)\right]\text{sgn}\left[\sin\left(\theta_{2}-\theta_{1}\right)\right] \right]
\end{align}
Finally, we just need to remember that as the trajectories that pass through both poles are topologically nonequivalent to those that do not, and notice that if we continuously change $\varphi$, the gap remains open (the spectrum is gapless only at points $\theta_1 \pm \theta_2 = n\pi$).
Hence, the different loops are deformed out of the plane breaking chiral symmetry, but remain topologically nonequivalent due to PHS and characterizing the two phases.
This is illustrated in Fig.~\ref{fig:CompositeTrajectories}. 
The red loop belongs to the phase in which the loop passes through both poles at $\varphi = 0$ and can be adiabatically deformed to a loop that again passes through both poles at $\varphi=\pi$. 
Similarly, the blue loop crosses a single pole twice at $\varphi = 0$ and at $\varphi =\pi$.
The topologically non-equivalence between the two types of loops makes it impossible for a loop crossing the two poles at $\varphi=0$ to become a loop that crosses twice the same pole at $\varphi=\pi$.
Note that the topology of these loops is not to be understood as the deformation of the two colored loops on the surface of a sphere (they would both be equivalent), but rather as a parity symmetry for the loops, produced by the existence of PHS.
Therefore, the invariant Eq.~(\ref{inv_supp}) appropriately predicts the existence of chiral edge states in the system.
%%%%%%%%%%%%%%%%%%%%%%%%%%%%%%%%%%%%%%%%%%%%%%%%%%%%%%%%%%%%%%%%%%
%%%%%%%%%%%%%%%%%%%%%%%%%%%%%%%%%%%%%%%%%%%%%%%%%%%%%%%%%%%%%%%%%%
%%%%%%%%%%%%%%%%%%%%%%%%%%%%%%%%%%%%%%%%%%%%%%%%%%%%%%%%%%%%%%%%%%

%%%%%%%%%%%%%%%%%%%%%%%%%%%%%%%%%%%

%%%%%%%%%%%%%%%%%%%%%%%%%%%%%%%%%%%

%%%%%%%%%%%%%%%%%%%%%%%%%%%%%%%%%%%%%%%%%%%%%%%%%%%%%%%%%%%%%%%%%%
%%%%%%%%%%%%%%%%%%%%%%%%%%%%%%%%%%%%%%%%%%%%%%%%%%%%%%%%%%%%%%%%%%
%%%%%%%%%%%%%%%%%%%%%%%%%%%%%%%%%%%%%%%%%%%%%%%%%%%%%%%%%%%%%%%%%%
\section{Analysis in terms of time-glide symmetry\label{appendix:Glide}}
Here we comment on the equivalence between our topological analysis in the main text and the one based on time-glide symmetry.

Time-glide symmetry is present in systems that are invariant under a time-translation plus a reflection. In particular, the glide operation acting on our Floquet operator, which is decomposed in 2-steps $U_\text{QW}(\mathbf{q})=U_2(\mathbf{q}) U_1(\mathbf{q})$, gives:
\begin{equation}
    \mathcal{G}_\text{T} U_\text{QW}(\mathbf{q}) \mathcal{G}_\text{T}^\dagger = \tilde{U}_\text{QW}(g\mathbf{q})
\end{equation}
where $g \mathbf{q}=(k,-\varphi)$ and $\tilde{U}_\text{QW}(g\mathbf{q})=U_1(\mathbf{q}) U_2(\mathbf{q})$ changes the order of the unitary matrices.
This means that for a system with this symmetry, we can write the following relation between the unitary matrices for each step:
\begin{equation}
    \mathcal{G}_\text{T} U_1(\mathbf{q}) \mathcal{G}_\text{T}^\dagger = U_2(g\mathbf{q})
\end{equation}
Consider the glide-symmetric plane $\mathbf{q}_\text{G}=(k,\varphi_\text{G})$, with $\varphi_\text{G}= \{ 0,\pi \}$. It is possible to define an effective Floquet Hamiltonian:
\begin{eqnarray}
    H_1(k)=i\log \mathcal{G}_\text{T} U_1(\mathbf{q}_\text{G})
\end{eqnarray}
which is twice the standard Floquet Hamiltonian and is gapped. Importantly, in the glide-symmetric plane $H_1(k)$ has chiral symmetry, and one can define its winding number, which coincides with the one obtained in the main text.
%%%%%%%%%%%%%%%%%%%%%%%%%%%%%%%%%%%%%%%%%%%%%%%%%%%%%%%%%%%%%%%%%%
\section{Extrinsic topology from the analogy with a dimers chain\label{appendix:AppendixSSH}}
As mentioned in the main text, the topology of our 2-step walk can be partially connected with the topology of a dimers chain. In particular, the Floquet operator for the 2-step walk takes a matrix form similar to that of the dimers chain, with unitary instead of Hermitian operators. Fortunately, this also provides a natural way to understand extrinsic topology. 

The Floquet operator can be written as:
\begin{equation}
    U_{F}=\left(\begin{array}{ccccccc}
U_{L} & U_{+} & 0 & 0 & 0 & 0\\
U_{-} & U_{0} & U_{+} & 0 & 0 & \ddots & 0\\
0 & U_{-} & U_{0} & U_{+} & \ddots & 0 & 0\\
0 & 0 & U_{-} & \ddots & U_{+} & 0 & 0\\
0 & 0 & \ddots & U_{-} & U_{0} & U_{+} & 0\\
0 & \ddots & 0 & 0 & U_{-} & U_{0} & U_{+}\\
 & 0 & 0 & 0 & 0 & U_{-} & U_{R}
\end{array}\right)
\end{equation}
where the different blocks are given by:
\begin{align}
    U_{0}=&\sin\left(\theta_{2}\right)\left(\begin{array}{cc}
-e^{-i\varphi}\sin\left(\theta_{1}\right) & ie^{-i\varphi}\cos\left(\theta_{1}\right)\\
ie^{i\varphi}\cos\left(\theta_{1}\right) & -e^{i\varphi}\sin\left(\theta_{1}\right)
\end{array}\right),\\
U_{+}=&\cos\left(\theta_{2}\right)\left(\begin{array}{cc}
0 & 0\\
i\sin\left(\theta_{1}\right) & \cos\left(\theta_{1}\right)
\end{array}\right),\\
U_{-}=&\cos\left(\theta_{2}\right)\left(\begin{array}{cc}
\cos\left(\theta_{1}\right) & i\sin\left(\theta_{1}\right)\\
0 & 0
\end{array}\right).
\end{align}
Notice also the presence of edge unitaries:
\begin{align}
    U_{L}&=\left(\begin{array}{cc}
-e^{-i\varphi}\sin\left(\theta_{1}\right) & ie^{-i\varphi}\cos\left(\theta_{1}\right)\\
ie^{i\varphi}\cos\left(\theta_{1}\right)\sin\left(\theta_{2}\right) & -e^{i\varphi}\sin\left(\theta_{1}\right)\sin\left(\theta_{2}\right)
\end{array}\right)\\U_{R}&=\left(\begin{array}{cc}
-e^{-i\varphi}\sin\left(\theta_{1}\right)\sin\left(\theta_{2}\right) & ie^{-i\varphi}\cos\left(\theta_{1}\right)\sin\left(\theta_{2}\right)\\
ie^{i\varphi}\cos\left(\theta_{1}\right) & -e^{i\varphi}\sin\left(\theta_{1}\right)
\end{array}\right)
\end{align}
which are different due to the condition of full reflection at the edges.
In this case, the matrix can be interpreted as a unitary version of a dimerized lattice, with couplings between neighboring dimers given by $U_{\pm}$.

As for the dimers chain, it is useful to check the different limits of dimerization, which in this case correspond to coupling only within each dimer ($\theta_1=0$ and $\theta_2=\pi/2$), or full hybridization between dimers ($\theta_1=\pi/2$ and $\theta_2=0$). In the first case the edge unitaries become:
\begin{equation}
U_{L,R}=\left(\begin{array}{cc}
0 & ie^{-i\varphi}\\
ie^{i\varphi} & 0
\end{array}\right)
\end{equation}
Their winding number can be calculated from~\cite{Bessho2022}:
\begin{equation}
    \nu_{edge}[U_\text{edge}(\varphi)] = \int_0^{2\pi}\frac{d\varphi}{2\pi}\text{tr}[U_\text{edge}(\varphi)^{-1} i \partial_\varphi U_\text{edge}(\varphi)],
\end{equation}
and results in a vanishing winding number.
In contrast, the edge unitaries for the second case result in:
\begin{equation}
    U_{L}=\left(\begin{array}{cc}
-e^{-i\varphi} & 0\\
0 & 0
\end{array}\right),\ U_{R}=\left(\begin{array}{cc}
0 & 0\\
0 & -e^{i\varphi}
\end{array}\right)
\end{equation}
and each edge unitary has non-vanishing winding number controlled by $\varphi$. Notice that they are completely decoupled from the bulk unitaries.
This exemplifies what is the meaning of edge unitaries in discretized time-step walks and how their topology (extrinsic) complements the bulk topology (intrinsic).

It is important to stress that this result explicitly shows that the topology of the 2-step walk is richer than that of a dimers chain. In particular, the fact that we are dealing with driven systems, characterized by unitary operators rather than Hermitian ones is what makes possible the existence of chiral edge states crossing both gaps.

Also, notice that as in this work we have characterized the topology of the system by linking the role of boundaries with different time frames and then, calculating the invariant for PBC.
This comparison with the dimers chain explicitly shows the edge unitaries and illustrate the connection with the  approach to extrinsic topology from Ref.~\cite{Bessho2022}, where a recipe to find the edge unitaries is not given.

\end{document}